\documentclass{optica-article}

\journal{opticajournal} 

\articletype{Research Article}

\usepackage{lineno}

\newcommand{\Visav}{$\langle \bar{V}\rangle_{BL}$}
\newcommand{\Errav}{$\langle \sigma_V \rangle_{BL}/\langle \bar{V}\rangle_{BL}$}
\newcommand{\Sigav}{$\langle \sigma_V \rangle_{BL}$}
\newcommand{\RP}{$\sigma_V / \bar{V}$}

\begin{document}

\title{Six-telescope integrated optics beam combiner fabricated using ultrafast laser inscription for J- and H-band astronomy}

\author{Aline N. Dinkelaker~\authormark{1}, Sebastian Smarzyk\authormark{1,2}, Abani S. Nayak~ \authormark{1,3}, Simone Piacentini~\authormark{4,5}, Giacomo Corrielli~\authormark{5}, Roberto Osellame~\authormark{5}, Ettore Pedretti~\authormark{6}, Martin M. Roth~\authormark{1} and Kalaga Madhav~\authormark{1}}

\address{\authormark{1}Leibniz-Institut f\"ur Astrophysik Potsdam (AIP), An der Sternwarte 16, 14482 Potsdam, Deutschland\\
\authormark{2}{Jade Hochschule, Friedrich-Paffrath-Str. 101, 26389 Wilhelmshaven, Deutschland}\\
\authormark{3}Institut f\"ur Angewandte Physik, Friedrich-Schiller-Universit\"at Jena, Albert-Einstein-Stra\ss e 15, 07745 Jena, Deutschland\\
\authormark{4}Dipartimento di Fisica - Politecnico di Milano, Piazza L. da Vinci 32 - 20133 Milano, Italia\\
\authormark{5}Istituto di Fotonica e Nanotecnologie (IFN) - CNR, Piazza L. da Vinci 32 - 20133 Milano, Italia\\
\authormark{6} UKRI STFC Rutherford Appleton Laboratory, Chilton, UK}



\begin{abstract*} 
We have built and characterized the first six-telescope discrete beam combiner (DBC) for stellar interferometry in the astronomical J-band. It is the DBC with the largest number of beam combinations and was manufactured using ultrafast laser inscription (ULI) in borosilicate glass, with a throughput of $\approx 56\%$. For calibration of the visibility-to-pixel matrix (V2PM), we use a two-input Michelson interferometer and extract the complex visibility. A visibility amplitude of 1.05 and relative precision of $2.9\%$ and $3.8\%$ are extracted for 1328~nm and 1380~nm, respectively. Broadband (< 40 nm) characterization is affected by dispersion, but shows similar performance. 

\end{abstract*}

\section{Introduction}

Long baseline stellar interferometry measures astronomical objects at highest angular resolution, enabling direct measurements of stellar diameters~\cite{Michelson:1921}, imaging stellar surfaces \cite{Monnier:2007,Roettenbacher:2016,Parks:2021}, exoplanets~\cite{LacourExo:2019}, protoplanetary structures around stars \cite{Kluska:2020}, or active galactic nuclei (AGN)~\cite{Hoenig:2018}, an overview can be found in e.g.~\cite{eisenhauer:23}. 
The Center for High Angular Resolution Astronomy (CHARA) Array~\cite{ten_Brummelaar:2005} and the Very Large Telescope Interferometer (VLTI)~\cite{vltiweb} are the current operating optical/infrared long-baseline interferometers that have baseline lengths of more than one hundred meters. The CHARA Array consists of six 1-m telescopes, while the VLTI consists of four telescopes (either the 8.2~m unit telescopes or the 1.8~m auxiliary telescopes). Near-infrared interferometry is commonly used, especially in the H (1450 - 1800~nm) and K (1950 - 2350~nm) -bands, but the J-band (1100 - 1400~nm) is relatively unexplored. It is particularly interesting for stellar astrophysics~\cite{Labdon:21, Anugu:2020, Pedretti:2018}, e.g. to access specific emission lines, for measurements of the photosphere, and to complement existing wavelength range measurements. 

 Astrophotonic components~\cite{Bland-Hawthorn:09,Norris:19,Dinkelaker:21,Jovanovic:2023} are increasingly used as alternatives to free-space optics in long baseline interferometry, thereby reducing the size and weight of an instrument with the potential to reduce sensitivity to environmental changes (e.g. temperature, mechanical vibrations). For N telescopes in an interferometric array, there are $\frac{N\cdot(N-1)}{2}$ resulting baselines and thus required beam combinations. Photonic beam combiners can provide the overlap for all inputs in the same cm-scale device. Not only that, with the small ($\mu$m-scale) and nearly freely configurable waveguide (WG) structures of chip- and fiber-based photonics, additional functionalities can be obtained that are hard or impossible to achieve with bulky free-space optics. Different types of integrated beam combiners have been developed and tested, e.g. the six-input (planar) SPICA-FT beam combiner currently under commission at the CHARA array~\cite{Pannetier:2022}, a recent hybrid device including a (planar) cascading beam combiner for aperture masking with the largest (28) number of baselines yet~\cite{Cvetojevic:21}, or the four-input (planar) GRAVITY beam combiner~\cite{Benisty:2009} at VLTI, contributing fascinating data about the motion of stars at the center of our galaxy~\cite{Gravity:2018}, thereby demonstrating maturity in performance and reliability of astrophotonics for their use in scientific instruments. To provide an overview, Tab.~\ref{tab:BCs} lists existing integrated optics beam combiners that are either part of interferometric instruments or have been subject to on-sky tests.

\begin{table}[hbt]
    \centering
    \begin{tabular}{l|c|c|c|c|c|c}
        \textbf{Name } & \textbf{Inputs} & \textbf{$\lambda$} & \textbf{Principle} & \textbf{Status} & \textbf{Facility} & \textbf{Ref.}\\

         \hline
         IONIC & 2 & H-band & Pairwise, temp. & O & VLTI & \cite{LeBouquin:2004}\\
         IONIC3 & 3  & H-band & Pairwise, temp. & C & IOTA & \cite{Berger:2003,Kraus:2005}\\
         PIONIER & 4 & H-band & Pairwise, matric. & C & VLTI & \cite{Benisty:2009, LeBouquin:2011} \\
         Dragonfly & 8 & H-band & All-in-one, spat. & O & AAT & \cite{Jovanovic:2012}\\
         GRAVITY & 4 & K-band & Pairwise, matric. & C & VLTI & \cite{Jocou:2014, GRAVITY:2017}\\
         GLINT & 4 & H-band & Pairwise, temp. & C & Subaru Tel. & \cite{Norris:2020, Martinod:2021}\\
         MYSTIC & 4 & K-band & Pairwise, matric. & UC & CHARA & \cite{Monnier:2018,Anugu:2020}\\
         DBC & 4 & H-band & All-in-one, matric. & O & WHT & \cite{Nayak:21}\\
         SPICA-FT & 6 & H-band & Pairwise, matric. & UC & CHARA & \cite{Anugu:2020,Pannetier:2022}\\
         6T-DBC & 6 & J-/H-band & All-in-one, matric. & L & - & This work \\
    \end{tabular}
    \caption{Overview of integrated optics beam combiners that have either been tested on-sky or are part of instruments. Given is the name of the beam combiner or instruments, the number of inputs for combination (either telescopes or sub-apertures in aperture masking), and the wavelength range ($\lambda$) in terms of astronomical band. The principle of operation is described in terms of beam combination (pairwise or all-in-one) and fringe encoding (temporal, spatial or matricial), see~\cite{Lebouquin:2004b}. Shown is also the status, which indicates whether an integrated optics beam combiner has been tested on-sky (O), the instrument is under commission (UC) or is commissioned (C), and the respective telescope facility. Although the beam combiner presented here has been tested in the laboratory (L) and not yet at a telescope, it is added for comparison. 
    While there are other integrated optics beam combiners with laboratory results, these  are not included in this table. The final column includes references.  }
    \label{tab:BCs}
\end{table}

Among the integrated beam combiners, the discrete beam combiner (DBC) is a specific concept based on evanescent coupling of light from several input sites within an array of WGs. DBCs are studied for their use in near-infrared interferometry as they can deliver sensitive observations with lower output channels and thus lower camera noise (readout and dark current shot noise), and their straight WG arrays enable easier and more flexible fabrication compared to other, curved beam combiner designs. Details on the principle of operation and different designs can be found e.g. in ~\cite{Minardi:2010,Minardi:2012,MinardiSPIE:2012,Minardi:2015,Diener:17, Saviauk:13} (see also~\cite{Nayak2022} for a complete description of a similar device for four inputs). 
So far, DBCs for up to four inputs have been developed for the visible, the H- and L-band~\cite{Saviauk:13,Nayak:21, Diener:17}, with laboratory results (visible, L and H-band) as well as first on-sky results of a four-input DBC (H-band) in an aperture masking experiment.  

Photonic beam combiners for more inputs would demonstrate the scalability of such devices, which might become increasingly important for future interferometric facilities with large arrays, for high density pupil remapping and aperture masking experiments, and for future astrophotonic-based instruments at existing facilities. Creating this technology for additional wavelengths extends the suitability of photonics for different instruments and science cases. One six-input integrated beam combiner is currently being commissioned to operate in the H-band~\cite{Pannetier:2022} at the CHARA Array. However, there are no J-band integrated optics beam combiners, see also~\ref{tab:BCs}. Here, we have manufactured a DBC for the J- and H-band, which is suitable for applications at the CHARA Array. The fabrication and characterization of the six-input DBC (6T-DBC) are the focus of this paper. The design and fabrication are described in section~\ref{DBCdevice}, the laboratory setup and methods for characterization in section~\ref{ExpSetup}, and the results with a focus on the J-band characterization in section~\ref{results}.

\section{DBC device}
\label{DBCdevice}

For all existing DBCs, the designs are based on three-dimensional (3D) WG arrays. In particular, mostly devices with two stacked WG layers have been fabricated and characterized, in which the WGs of the DBC are arranged in a zig-zag geometry in the transverse plane (see Fig.~\ref{fig:scheme} for visualization). 

Light is injected at N input WGs, which are transported to an array of regularly and closely-spaced WGs, where the number of WGs, M, has to be $\geq N^2$. Light distributes from the input WGs across the array through evanescent coupling, which allows simultaneous beam combination of all inputs. How the light is distributed within the array depends on the location of the input WGs and the coupling strength to neighboring WGs, which determines the evolution of the field amplitude and phase along the length of the array, thus containing the complex visibility. The interference signal is obtained by measuring the light intensity at the M output WGs. An advantage of the DBC is that it acts as a static beam combiner, i.e. the complex visibility of the source can be measured without any temporal phase delay between the input arms of the DBC. Another feature of the DBC is that it uses an array of straight WGs, with no bends in the beam combiner itself, thus removing bending losses. Due to the simple WG configuration, DBCs offer a straight forward approach for scaling up to larger number of input beams. DBCs can also be combined with pupil remappers~\cite{Jovanovic:2012} or reformatters~\cite{Harris:2018} at the in- or output, which guide the light from a convenient location to or from the WG array using paths, however, that include bend structures. 

The 3D arrays of the DBCs --as well as additional WG structures for input and output reformatting-- have been fabricated using ultrafast laser inscription (ULI)~\cite{Thomson:09}. With this method, a femtosecond laser is focused inside a glass substrate, where thermal effects locally change the refractive index of the material. By moving the substrate relative to the focus, WG structures can be written into the substrate at different heights. With the ULI method, the extra dimension allows more degrees of freedom in the design of the DBC, e.g. the bend radii of WGs feeding the beam combiner section and the distance between WGs can be controlled in 3D. Various parameters have to be precisely controlled to produce smooth, single-mode WGs with ULI fabrication. The design and fabrication are described in more detail in the following sections. 

\subsection{Design of the 6T-DBC}
\begin{figure}[ht]
\centering\includegraphics[width=\textwidth]{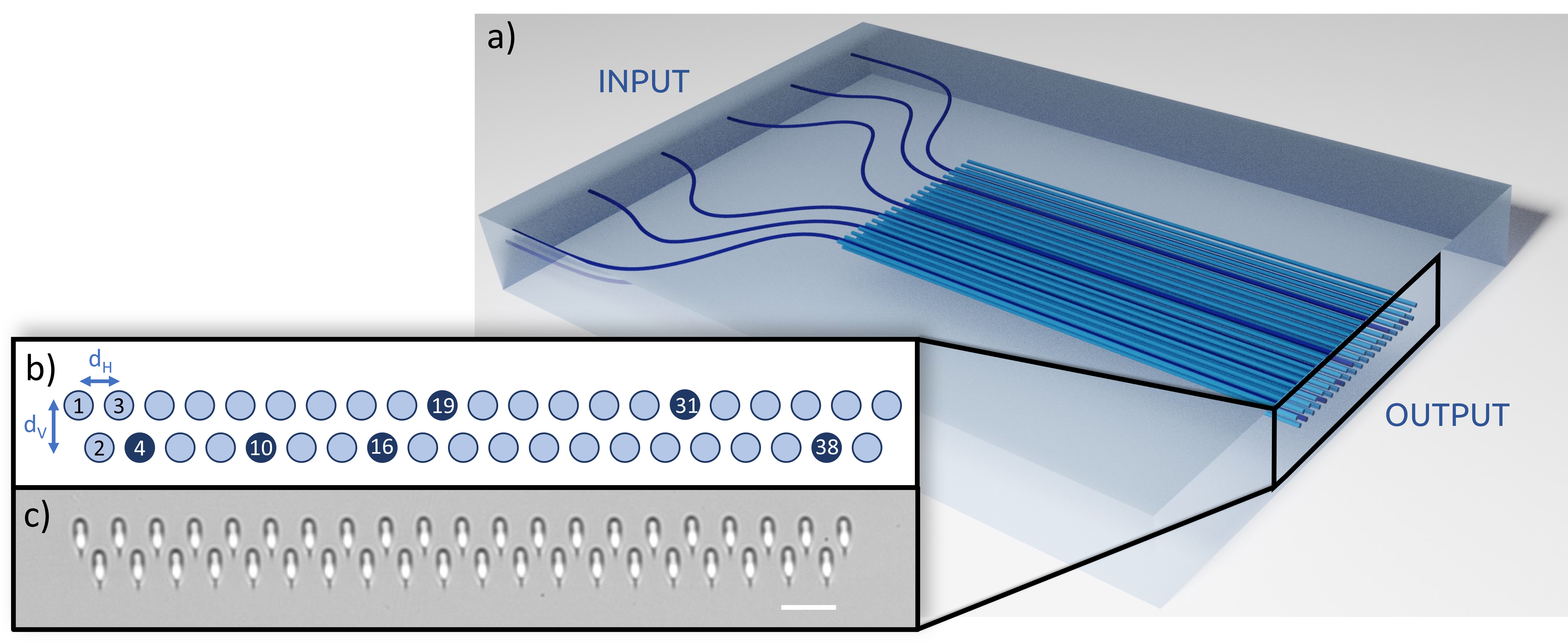}
\caption{a) Design of the zig-zag 6T-DBC with fan-in. The input WGs, in dark blue, are spaced by 127 \textmu m at the input end. The six WGs leading to the DBC region are curved so as to have the same path lengths. b) Schematic representation of the DBC section, where $d_H$ is the planar distance between the WGs and $d_V$ is the separation between the planes. The inputs are labeled as 4, 10, 16, 19, 31 and 38. c) Microscope image of the output section of a DBC with $d_H=16$ \textmu m and $d_V=13.3$ \textmu m. The scale bar indicates 20 \textmu m.}
\label{fig:scheme}
\end{figure}

We create devices with two different design types, at the core of which are DBCs with the same pattern. The fabricated DBCs are composed of 41 WGs, placed in a zig-zag configuration and spaced at a distance that enables evanescent coupling. This arrangement allows to independently control the planar and the diagonal coupling coefficients, since the former can be tuned by changing the planar separation of the WGs $d_H$, while the latter is dependent on the separation $d_V$ between the two planes (Fig.~\ref{fig:scheme}). The two designs differ in the specification of the input region: (1) in the proof-of-concept design, light is coupled into six straight input WGs leading to and continuing into the WG array, and (2) in the fan-in design, light can be injected in the array by means of a fan-in reformatting region composed of six WGs with collinear input ports spaced 127~\textmu m from each other in order to provide an optimal alignment with standard fiber arrays and to prevent evanescent coupling. These WGs, designed with circular arcs, have the same length for maintaining the optical path difference among the injected modes, see Fig.~\ref{fig:scheme}. After optimization, a radius of curvature of 60 mm was chosen as a compromise between low bending loss ($\ll 0.1$ dB/cm) and a compact footprint. 

\subsection{Optimization of the manufacturing process}
The DBC devices presented here were manufactured at Politecnico di Milano and Istituto di Fotonica e Nanotecnologie, Italy, using the ULI method. The WGs were inscribed in a borosilicate substrate (Eagle XG, Corning) by focusing the laser beam generated by a homemade Yb:KYW source (wavelength of 1030~nm, repetition rate of 1 MHz, pulse duration of 300~fs) with a microscope objective (50x magnification, 0.6~NA). The glass sample was translated at 40~mm/s by an air bearing three-axis motion stage (Aerotech FiberGlide 3D), which provides a resolution of 1~nm and a position repeatability of 50~nm. The energy per pulse was 440~nJ, and the translation process was performed five times for every WG to induce the required index contrast. After the inscription, a thermal annealing on the sample was performed, to reduce the WGs' birefringence and therefore achieve polarization insensitive operation \cite{corrielli2018}. The annealing consisted of two subsequent heating ramps, the first at a rate of 100~\textdegree C/h up to 600~\textdegree C, and the second up to 750~\textdegree C at a rate of 75~\textdegree C/h, followed by a cooling to room temperature at a rate of -12~\textdegree C/h. The complete fabrication process allowed to obtain single mode operation around the wavelength of 1310~nm, with propagation losses below 0.3~dB/cm and an average 1/e\textsuperscript{2} mode dimension of 6.5(3)~\textmu m x 7.0(3)~\textmu m. 

\begin{figure}[ht]
\centering\includegraphics[width=\textwidth]{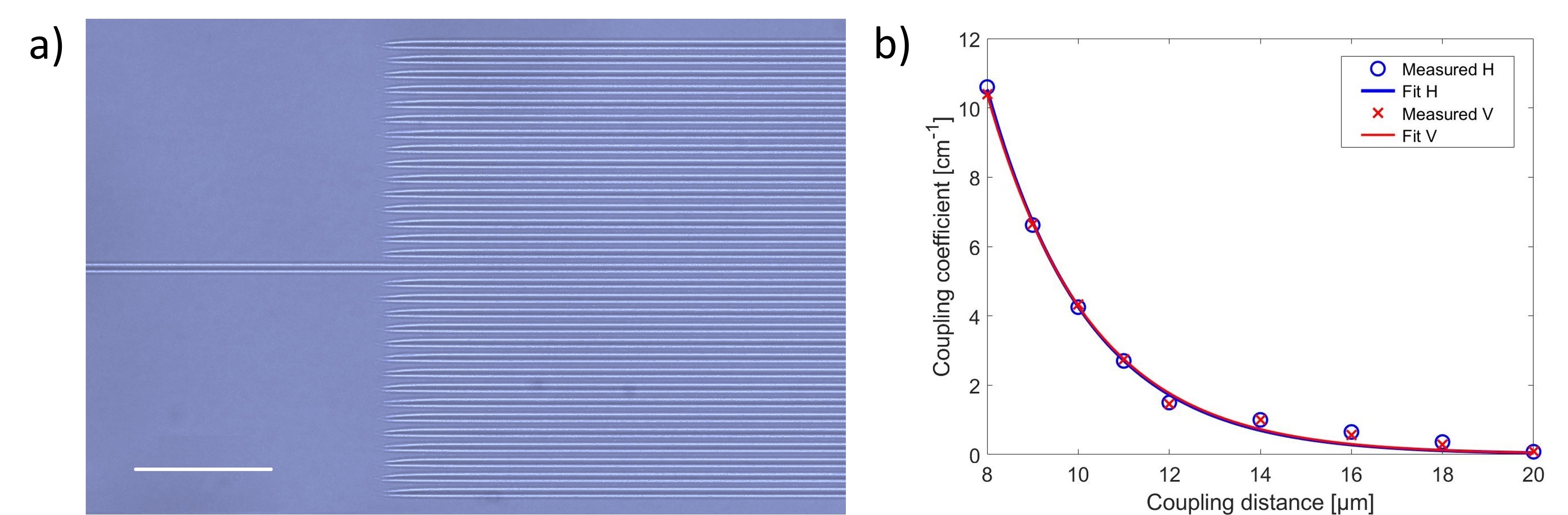}
\caption{a) Microscope image of a planar array with distance $d_H = 16$ \textmu m. The central WG is used to inject light in the array. The scale bar indicates 100 \textmu m. b) Exponential dependence of the planar coupling coefficient $k_H$ on the WG separation $d_H$. In blue the data related to horizontal (H) polarization are shown, while the vertical (V) polarization is represented in red. The continuous curve is retrieved by an exponential fit of the experimental data.}
\label{fig:planar}
\end{figure}

The dependence of the planar coupling coefficient $k_H$ on the WG  separation was studied by fabricating a set of linear arrays (Fig.~\ref{fig:planar}.a) with different interaction distances $d_H$, ranging from 8~\textmu m to 20~\textmu m. The arrays were characterized by injecting light at 1310~nm in the central WG, and by collecting the output intensity distributions with an IR camera. The measurement was performed for both horizontally and vertically polarized light, to characterize the polarization sensitivity of the fabricated WGs. By a fitting procedure, it was possible to retrieve from the collected images the coupling coefficient for different distances. This information is represented in Fig.~\ref{fig:planar}.b. As expected, the data show an exponential 
trend (according to the law $k_H=\alpha\cdot \exp\,(-\beta\cdot d_H)$, with $\alpha\simeq 340$~cm\textsuperscript{-1} and $\beta\simeq 0.44$~\textmu m\textsuperscript{-1}) and confirm the coupling polarization independence. 

\subsection{Initial characterization and DBC selection}
\label{DBCselection}
For both designs, DBCs with different separations between the planes $d_V$ are written onto the same chip, from which the one with the most suitable performance can be chosen. For the straight input DBCs, arrays with six different values for the vertical distance $d_V$ were manufactured (8.0, 9.5, 11.4, 13.8, 17.2, 22.0 ~$\mu$m, corresponding to angles between planes of 45$^\circ$, 50$^\circ$, 55$^\circ$, 60$^\circ$, 65$^\circ$, 70$^\circ$, respectively) with a constant array length $L = 40$~mm. The substrate has a footprint of $7 \times 50$~mm. For the fan-in DBCs, arrays with three different values for the WG array length $L$ (40, 45, 50 mm) were manufactured, each with four values for the vertical distance $d_V$ (10.2, 11.4, 12.8, 13.8~$\mu$m, corresponding to angles of 52$^\circ$, 55$^\circ$, 58$^\circ$, 60$^\circ$, respectively), and a chip footprint of $20 \times 71$~mm.

\begin{figure}[th]
\centering\includegraphics[width=0.8\textwidth]{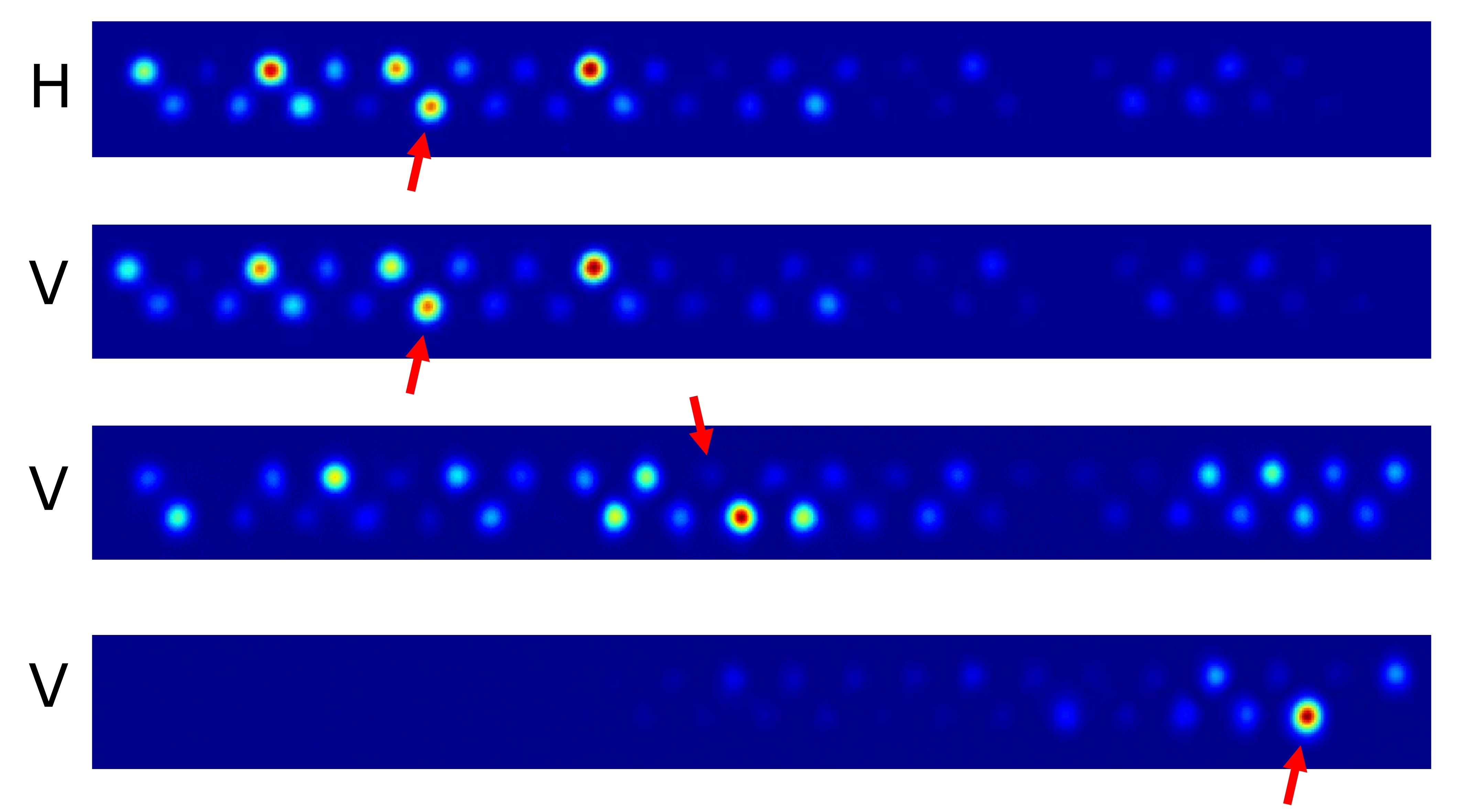}
\caption{Output intensity distribution of the fan-in DBC with $d_V = 11.4$ \textmu m and $L = 40$ mm when horizontally (H) or vertically (V) polarized light is injected in WG 10, indicated by the red arrow (top two images). The two distributions differ by less than $1\%$. Shown below are the distributions for (vertically polarized) light injected in WG~19 and in WG~38, marked with the red arrows. How the light spreads depends on the input sites and coupling properties. A precise and reliable estimation of the two coupling coefficients is not trivial due to the large number of WGs in the array.}
\label{fig:polarization}
\end{figure}

The planar separation $d_H$ was instead maintained constant at 16~\textmu m for all arrays. At this distance, the modified regions are well separated, thus making the fabrication more reproducible, and the mode overlap is non-negligible, with an expected coupling coefficient of $\kappa = 1$~cm\textsuperscript{-1}. From this value, $\frac{L}{L_c}$ can be calculated for a given wavelength  (we approximate $L_C \approx \pi/2\kappa $). The characterization of these devices with linearly polarized light confirmed the polarization independence of both the planar and the diagonal coupling. As an example, we report in Fig.~\ref{fig:polarization} the output intensity distribution of the fan-in DBC with $d_V = 11.4$ \textmu m and $L = 40$ mm when injecting H and V polarized light in WG~10. Also included are examples for light injected at WG~19 and at WG~38, which show the dependence of light distribution on the input site for this particular WG array. Moreover, by coupling light by an SMF28 fiber in the inputs of the DBCs, we measured an average insertion loss of $2.5\pm 0.3$ dB at 1310~nm, corresponding to a transmission of $56\%$.

\begin{figure}[th]
\centering\includegraphics[width=0.9\textwidth]{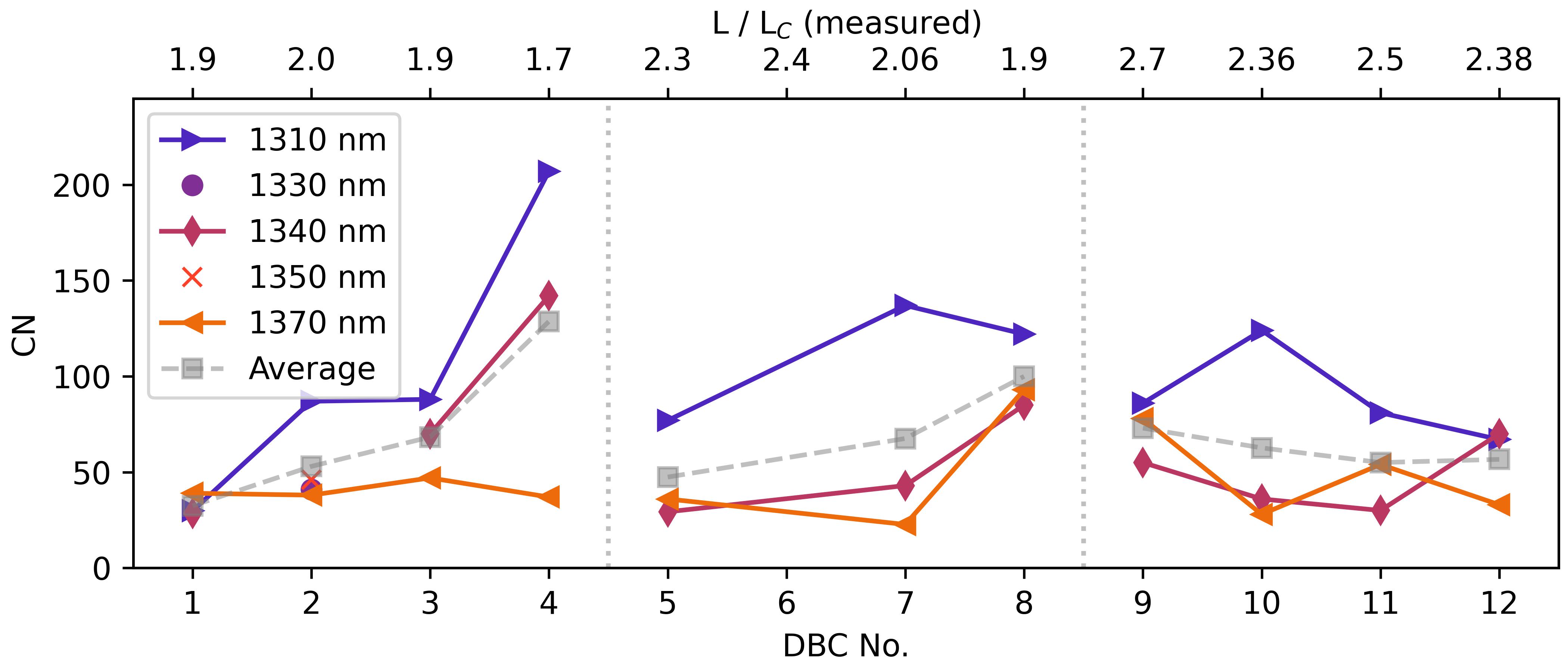}
\caption{Initial characterization for DBC selection: All 12~DBCs with fan-in (type~2) on the same chip have been characterized using monochromatic light at selected wavelengths: 1310~nm, 1340~nm and 1370~nm (different wavelengths were only used for DBC~2). Shown in the plot is the condition number (CN, see text for details) for each wavelength and the average, which was one of the main factors for DBC selection. We can divide the DBCs into three sets: DBCs 1~to~4 have $L = 40$~mm, DBCs 5~to~8 have $L = 45$~mm, and DBCs 9~to~12 have $L = 50$~mm, where $d_V$ increases with DBC number in each set. The top x-axis shows the corresponding measured $\frac{L}{L_c}$. Several DBCs have comparable performance, and DBC~5 was chosen for further characterization. Note that due to very high CN values and irregular output patterns, we suspected DBC~6 to have a fault. It was quickly removed from consideration and is not shown in this overview.}
\label{fig:DBCselection}
\end{figure}

In the next step, interferometric measurements (see the following sections for more detail on the measurement setup and procedure) at different wavelengths in the J-band were performed to identify good candidate DBC-devices for interferometry. DBCs that fulfill these criteria have a suitable combination of horizontal and vertical coupling strength and array length to enable light distribution from the input WGs throughout the array and obtain strong enough interferometry signals at the outputs without suffering from unnecessary propagation losses due to excessive array length. A subset of DBCs on the same chip gave similar results in term of interferometric signal quality, visibility, and stability, with condition numbers (CN, see section~\ref{CNsectionmethod}) comparable to simulations for the expected value of $\frac{L}{L_c}$. Figure~\ref{fig:DBCselection} shows the CN for the different fan-in (type~2) DBCs under consideration, which was one of the characterization results used for comparison.

The DBCs selected for this paper have the following properties: for the straight input design (type 1), the array length is $L = 40$~mm and the plane separation is $d_V = 8.02~\mu$m. For this DBC, the coupling constant is estimated from fits to output intensity measurements, with an average value of $\kappa = 1.4 \pm 0.4$~cm$^{-1}$ (the given uncertainty corresponds to the standard deviation from measurements at different input sites), but with stronger coupling in the diagonal ($\kappa = 1.8 \pm 0.2$~cm$^{-1}$) than the horizontal ($\kappa = 1.1 \pm 0.1$~cm$^{-1}$) direction. For the fan-in design (type 2), $L = 45$~mm and $d_V = 10.2~\mu$m. From the intensity distribution at the WG outputs, the coupling coefficient for these separations $d_H$ and $d_V$ is fitted as $\kappa = 0.8 \pm 0.2 $~cm$^{-1}$ in the diagonal as well as the horizontal direction. Due to fabrication variability, this is lower than the expected value of $1$~cm$^{-1}$, see Fig.~\ref{fig:planar}. Nevertheless, the results will be presented, since the fan-in configuration is a useful feature and the coupling coefficient can be increased for future devices. For comparison, Table~\ref{tab:DBCABCD} shows properties of the ULI-fabricated 6T-DBC as well as the integrated optics six-input SPICA-FT beam combiner~\cite{Pannetier:2022}.

\begin{table}[htb]
    \centering
    \begin{tabular}{l|l|l}
                 & \textbf{SPICA-FT} & \textbf{fan-in DBC} \\
         \hline
       Beam combiner type  & ABCD & DBC \\   
       Number of inputs  & 6 & 6 \\
       Number of outputs & 60 & 41 \\
       Operating wavelengths  & H-band  & J-band (/ H-band) \\
       Throughput & > 55\% & 56 \%\\
       Chip footprint  & $82~\text{mm} \times 35~\text{mm}$& $71~\text{mm} \times 20~\text{mm}$\\
       Fabrication  &  Lithography & ULI \\
       Instrumental contrast & $74\% \pm 13\%$  & $89\% \pm 7\%$\\
       Crosstalk & $0.1 \%$ & N/A \\
    \end{tabular}
    \caption{Side-by-side comparison of the six-input SPICA-FT beam combiner~\cite{Pannetier:2022} and the six-input DBC with fan-in. The instrumental contrast for SPICA-FT is given as the average for all baselines and 27 spectral channels between $1.45~\mu$m and $1.65~\mu$m~\cite{Pannetier:2022}. The instrumental contrast for the fan-in DBC is the average over all baselines for 51 monochromatic measurements between $1.28~\mu$m and $1.38~\mu$m, calculated from the V2PM as described in~\cite{Diener:17}. Crosstalk is not defined for a DBC, as light is expected to spread across all WGs. Note that the footprint of the fan-in DBC is for a chip with 12 beam combiners and could be reduced. }
    \label{tab:DBCABCD}
\end{table}

\section{Beam combiner test bench and method}
\label{ExpSetup}
\begin{figure}
\centering\includegraphics[width = 0.7 \textwidth]{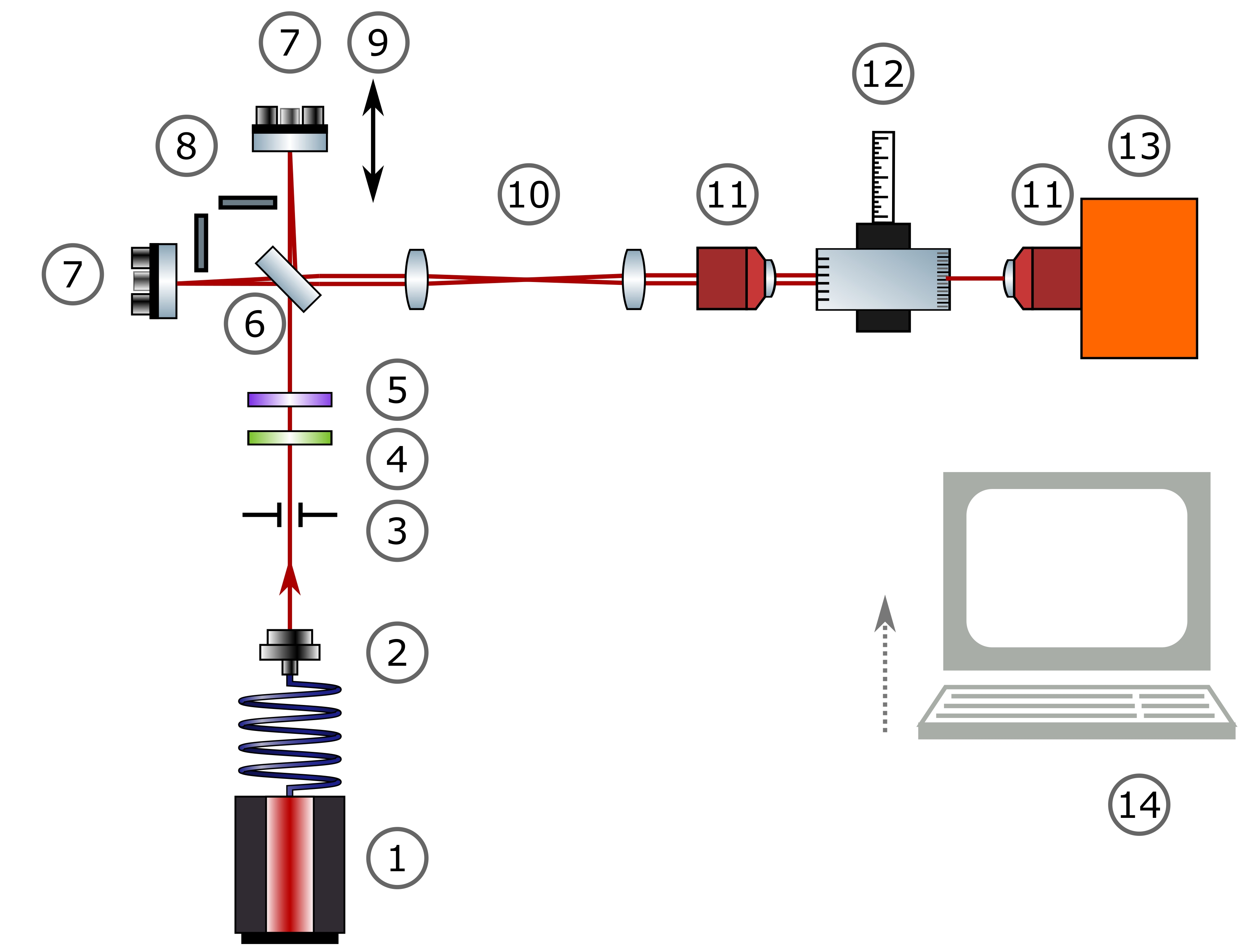}
\caption{Schematic of the interferometric setup for characterization (not to scale). Light from the fiber-coupled source (1) enters the setup through a fiber collimator (2) and passes an iris (3), neutral density filters (4) and a polarizer (5). The light is then split using a 50:50 beam splitter (6) and forms a Michelson interferometer. The mirrors in both arms (7) can be steered using motorized actuators and blocked with motorized shutters (8), one mirror is mounted on a motorized translation stage (9) for optical path delay. The beams then pass beam reducing optics (10) and coupling optics (11). Light is coupled into the photonic beam combiner (12) vacuum mounted on a 5-axis positioner. The output light is recorded with a camera (13). A control computer (14) is used for automated measurements.}
\label{fig:expsetup}
\end{figure}

A Michelson-type setup is used to characterize the DBCs interferometrically using two beams with variable optical path difference (OPD). The experimental setup has been described in \cite{Pedretti:2018}, with a few changes for additional automation. The setup is shown schematically in Fig.~\ref{fig:expsetup}. A fiber-coupled computer-controlled light source feeds a fiber polarization controller (not shown) and a collimator. The beam size is adjusted with an iris. Neutral density filters and a polarization plate are used for fine control of the laser intensity. Using a 50:50 beam splitter, the beam is split and back reflected from two gold-coated mirrors on motorized kinematic mounts (TRA12PPD actuators controlled with SMC100CC) in this Michelson-type setup. After alignment, however, the two beams are displaced relative to each other to feed different inputs and the interferometric overlap happens inside the chip. The optical path of one beam is delayed with respect to the other using a motorized translation stage (MTS25-Z8 controlled with KDC101). Light in each arm can be blocked using motorized shutters. The two beams pass through a beam reducer after which they reach the coupling optics. After the chip, following the output coupling optics for magnification, a camera records images of the chip's output facet (Raptor Photonics Ninox 640, cooled to -6$^{\circ}$ with 5.8~ms exposure time for the majority of the data).

Initial alignment of the integrated-optics component to the centered, horizontally flat and overlapped beams from the interferometer is done using a 5-axis positioner (Luminos i5000), with which light is coupled into the photonic beam combiner under characterization. The two beams can then be independently steered in order to address two independent inputs of the beam comber using the motorized kinematic mounts. The inputs of the component are found by raster scanning the mirror mount stepper motors and building a map of the intensity of the outputs of the component as a function of the \(x,y\) position of each mirror. The six inputs are identified on the map and their individual positions saved, allowing to address any pairwise combination of inputs to perform measurements for all baselines. The movable mirrors, delay line, shutters and camera are all controlled via software and synchronized for automated measurements. 

The characterization of each DBC can be divided into two parts: First, a fiber coupled light source is used as an ideal point source --with an expected visibility amplitude of 1-- to calibrate the DBC's transfer matrix, the visibility-to-pixel-matrix (V2PM), and estimate its stability. Second, the V2PM is inverted to obtain the pixel-to-visibility-matrix (P2VM) and used to extract the visibility of the light source to verify that the expected amplitude value of 1 can be retrieved. Note that the value for the extracted visibility does not provide information about the instrumental contrast, as it is self-calibrated~\cite{Tatulli:2007}, but an estimate of the instrumental contrast can be calculated from the V2PM, see~\cite{Diener:17}. The two steps, calibration and verification, are performed for different monochromatic and broadband (bandwidth < 40~nm) light sources. 
For the results presented here, the V2PM is populated by injecting light at two inputs at a time (pairwise) using the method shown in~\cite{Saviauk:13}, which is based on the description and procedure detailed e.g. in~\cite{Tatulli:2007} and~\cite{Lacour:2008}. We have $N = 6$ inputs, thus the V2PM has $N^2 = 36$ columns. The number of rows corresponds to the number of output WGs, here $M = 41$. The important steps for the calibration and verification process will be described in the following section in brief, further details can be found in~\cite{Saviauk:13}.

\subsection{V2PM calibration}

The V2PM relates the coherence vector $\vec{J}$ to the intensity $\vec{I}$ measured at the $M$ output WGs of the DBC:
\begin{equation}
    \vec{I} = V2PM \cdot \vec{J},
\end{equation}
which corresponds to a system of linear equations and can also be displayed as: 
\begin{equation}
    I_n = \sum_{k=1}^{N^2 = 36}\alpha_{nk}J_k,
\end{equation}
where $n=1,\dots,M = 41$ (see Eq.~1 in \cite{Saviauk:13}). With the V2PM used here (the $\alpha$-matrix described in~\cite{Minardi:2012,Saviauk:13}), the coherence vector $\vec{J}$ consists of two parts: first are $N = 6$ photometric terms, the self-coherence $\Gamma_{ii}$ of a beam in each input $i$, corresponding to the input intensities. These are followed by $N\cdot(N-1)$ interferometric terms, including the real and the imaginary parts of the mutual coherence $\Gamma_{ij}$, where $i$ and $j \neq i$ are the inputs:

\begin{equation}
    \vec{J} = (\Gamma_{11},\dots,\Gamma_{NN},\mathfrak{Re}\Gamma_{12},\mathfrak{Im}\Gamma_{12},\dots,\mathfrak{Re}\Gamma_{(N-1)N},\mathfrak{Im}\Gamma_{(N-1)N}).
    \label{eq:gamma}
\end{equation}

The order in which the V2PM is arranged determines the order of the resulting coherence vector when multiplying its inverse with the intensity measurements. From the coherence vector, the complex visibility can be calculated (Eq.~3 and 4 in~\cite{Saviauk:13}). 

The V2PM is experimentally determined through calibration measurements. For all 15 combinations, the interferometric signal is recorded by steering the motorized mirrors such that the selected input sites are addressed, opening both shutters, and scanning the delay line in one arm over a few ($\sim 10$) $\mu$m (the OPD will be twice that due to the mirror reflection). The photometries are recorded for each input by closing one of the shutters, leading to a total of 45 files for the 15 combinations. The number of samples (or frames) corresponds to the steps of the delay line position. The data is saved as FITS-files, including the camera image section of ca. 50 x 600 pixel (exact size can be adjusted) for all samples (ca. 300 for monochromatic data). An automatically generated filename from the names of the input sites and the type of measurement allows its use as an identifier for the data analysis. 
\subsubsection{Data reduction}
The FITS-files are read in and data reduction and analysis are performed in Python. As a first step, the (central) coordinates of the 41 output waveguides have to be found, for which all images of the same measurement are combined. The outputs can be detected with a pattern recognition function. However, with a regular zig-zag WG arrangement at the output, a more stable hybrid solution is used: the x and y coordinates of the first WG as well as the spacing in x and y (and a linear tilt) between WGs are stored in a json file. These parameters are roughly, manually adjusted by visually comparing the image with the coordinates marked on top. In a second step, the maximum around these coordinates is found automatically. Finally, all (x,y)-coordinates within a circle of a specified radius around the center are selected to form the full set of coordinates for which the data is extracted from the images. In an ideal scenario, intensity variations at the WG outputs could be monitored with a single pixel at each output, or even photodiode arrays, to minimize the number of pixels and thus the detector noise. In our test bench, this is not optimized, and each WG output is imaged onto more than one pixel on the detector. We therefore increase the number of pixels to capture more signal and to mitigate any inaccuracy in the semi-automated output search algorithm. For the data in this paper, 12 pixels per WG contribute to the data selected for analysis.

\subsubsection{Data fitting to populate the V2PM}

To populate the V2PM, several parameters have to be obtained from the calibration data. The first N columns of the V2PM are filled with the transmission coefficients, $\alpha_{nk} = \alpha_{nii}$ ($ k = i = 1,\dots,N$), calculated from the fraction of intensity at the $n$-th output WG relative to the total intensity for a given (the $i$-th) input:

\begin{equation}
\alpha_{nii}=\frac{I_{nii}}{\sum_n^{41} I_{nii}}.   
\label{eq:transmissioncoeff}
\end{equation}

With 15 baselines, we have 5 recorded photometries per input, from which the average transmission coefficients are calculated for each input. This method does not account for losses within the device. The remaining $N\cdot(N-1)$ columns of the V2PM are filled using the results of fitted functions to the recorded interference fringes, following the procedure and Eq.~7-12 in~\cite{Saviauk:13}, where more details can be found. With light at inputs~$i,j$, we record the signal $I_{nij}$ at the $n$-th output WG. The static (DC) component is removed from the signal $I_{nij}$ to get the oscillating (AC) term by subtracting the corresponding measured photometry (here, we use a 3rd order polynomial fit to the photometry data):

\begin{equation}
    \tilde{I}_{nij} =  I_{nij} - I_{nii} - I_{njj}.
        \label{eq:FringeAC}
\end{equation}

Note that for our data analysis, we divide the interferometric term by the input field amplitudes before fitting, where we use the recorded photometries to obtain the input intensities (ignoring losses), $\Gamma_{ii} = \sum_n^{41} I_{nii}$, see also~Eq.~\ref{eq:transmissioncoeff}. This step only pulls forward the normalization for the calculation of the visibility amplitude, see Eq.~\ref{eq:vis}, and we obtain: 
\begin{equation}
    \tilde{I}_{nij\ \textrm{norm}} =  \frac{I_{nij} - I_{nii} - I_{njj}}{\sqrt{\Gamma_{ii} \Gamma_{jj}}}.
        \label{eq:Fringenorm}
\end{equation}

The normalized interferometric term can be fitted using
\begin{equation}
    \tilde{I}_{nij\ \textrm{norm}} =  a_{nij} \cos({x \cdot k}) + b_{nij} \sin({x \cdot k}) + c,
    \label{eq:SinCos}
\end{equation}
where x is the delay line position that changes the optical path, k the corresponding (spatial) frequency, and c is an offset. For a given baseline with inputs $i,j$, the fitted amplitudes $a_{nij}$ and $b_{nij}$ (generalized: $a$ and $b$) fill the two interferometric columns of the V2PM, which connect to the real and imaginary part of the mutual coherence $\Gamma_{ij}$ (Eq.~\ref{eq:gamma}), respectively (see~\cite{Saviauk:13}).

\begin{figure}[ht]
\centering \includegraphics[width=.7\textwidth]{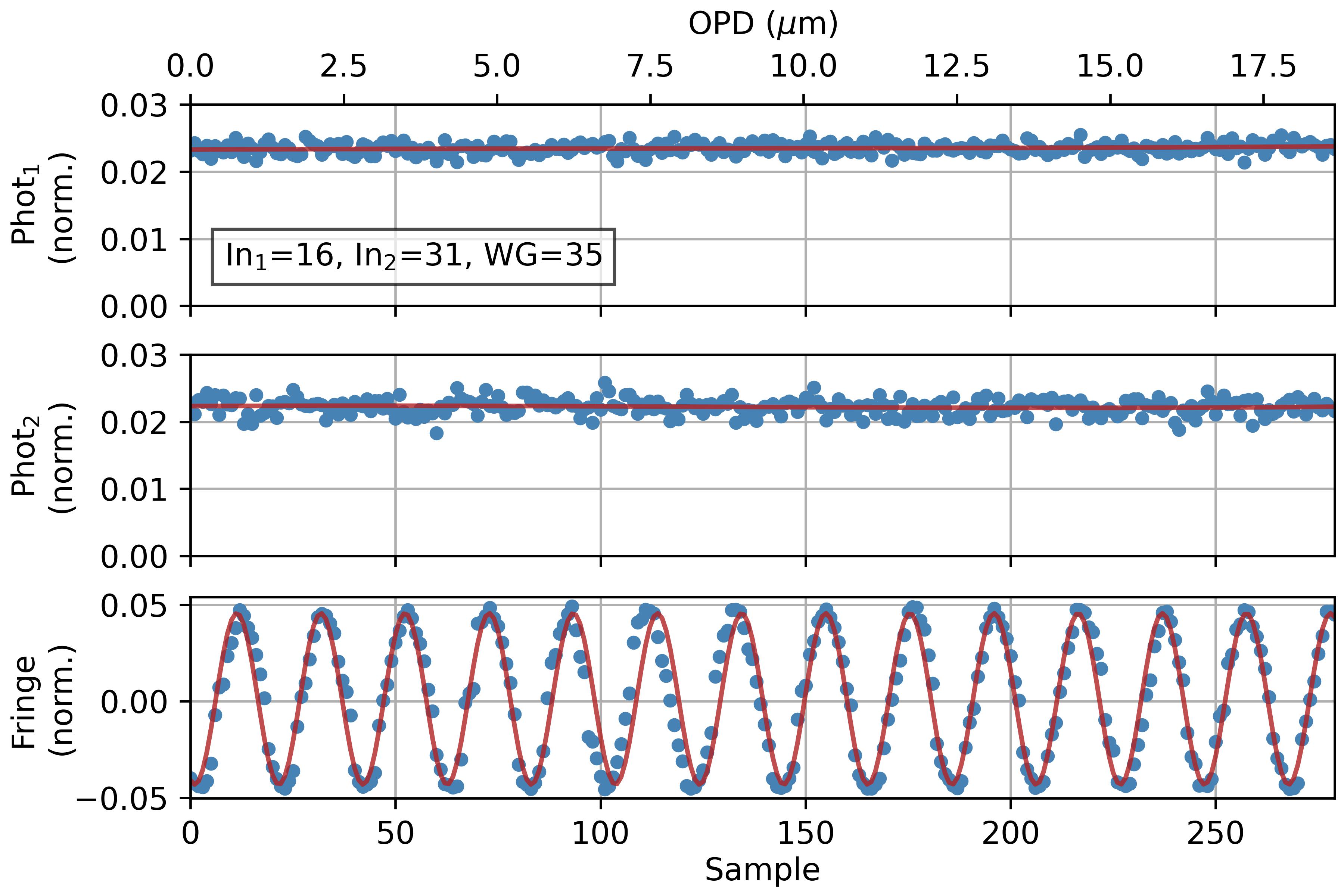}
\caption{Top and middle: Example of the two input photometries (normalized to give the relative transmission $\alpha_{nii}$) and the fit that is used for matrix calibration. Bottom: Corresponding interferometric fringe signal, namely the normalized intensity $\tilde{I}_{nij\ \textrm{norm}}$ from Eq.~\ref{eq:Fringenorm} and its fit using Eq.~\ref{eq:SinCos}. The plots show monochromatic measurements at 1380~nm for the same output WG. The x-axis is the same for all plots, given are the sample number (bottom) as well as the relative OPD (top) caused by the delay line movement. The y-axis is normalized and thus unitless.}
\label{fig:fringefitexample}
\end{figure}

We assume that during a measurement with two fixed input sites, the frequency is constant for the fringes at all output WGs and thus apply a multi-step fitting procedure. First, the initial values for the variable parameters are found from the power spectrum after performing a fast Fourier transform (FFT). With these starting values, a fit is performed of Eq.~\ref{eq:SinCos} to the interferometry data at each output WG. From the first round of fit results, k$_{median}$, the median of the WGs' fitted frequency, is calculated (to avoid outliers due to low signal-to-noise in individual WGs). Next, the fitting procedure is repeated, but with restricted frequency variation within k$_{median} \pm 30\%$. From these results, the mean of the WGs' fitted frequencies is calculated and used as fixed value for k in the final fitting round, leading to a consistent frequency for all WGs with appropriately fitted amplitudes $a$ and $b$. Examples of photometric and interferometric data with fitted functions are shown in Fig.~\ref{fig:fringefitexample}. The process is repeated for each baseline.

The V2PM columns $N + 1$ to $N^2$ are filled using the fit results for $a$ (columns $N+1, N + 3,\dots,N^2-1$) and $b$ (columns $N+2, N+4, \dots, N^2$). With this, the columnwise calibration of the V2PM is concluded and its pseudo-inverse as well as the condition number (see section~\ref{CNsectionmethod}) are calculated. 
With wavelength-dependent coupling, the calibration has to be performed for each wavelength separately. Note that in our case, the numbers of the input sites correspond to their position in the array, which results in the mapping $(1,2,3,4,5,6) \rightarrow (4,10,16,19,31,38)$. This numbering will be used to identify the input sites throughout the paper. 

\subsubsection{Stability of the V2PM - the condition number}
\label{CNsectionmethod}
\begin{figure}
\centering \includegraphics[width=0.95 \textwidth]{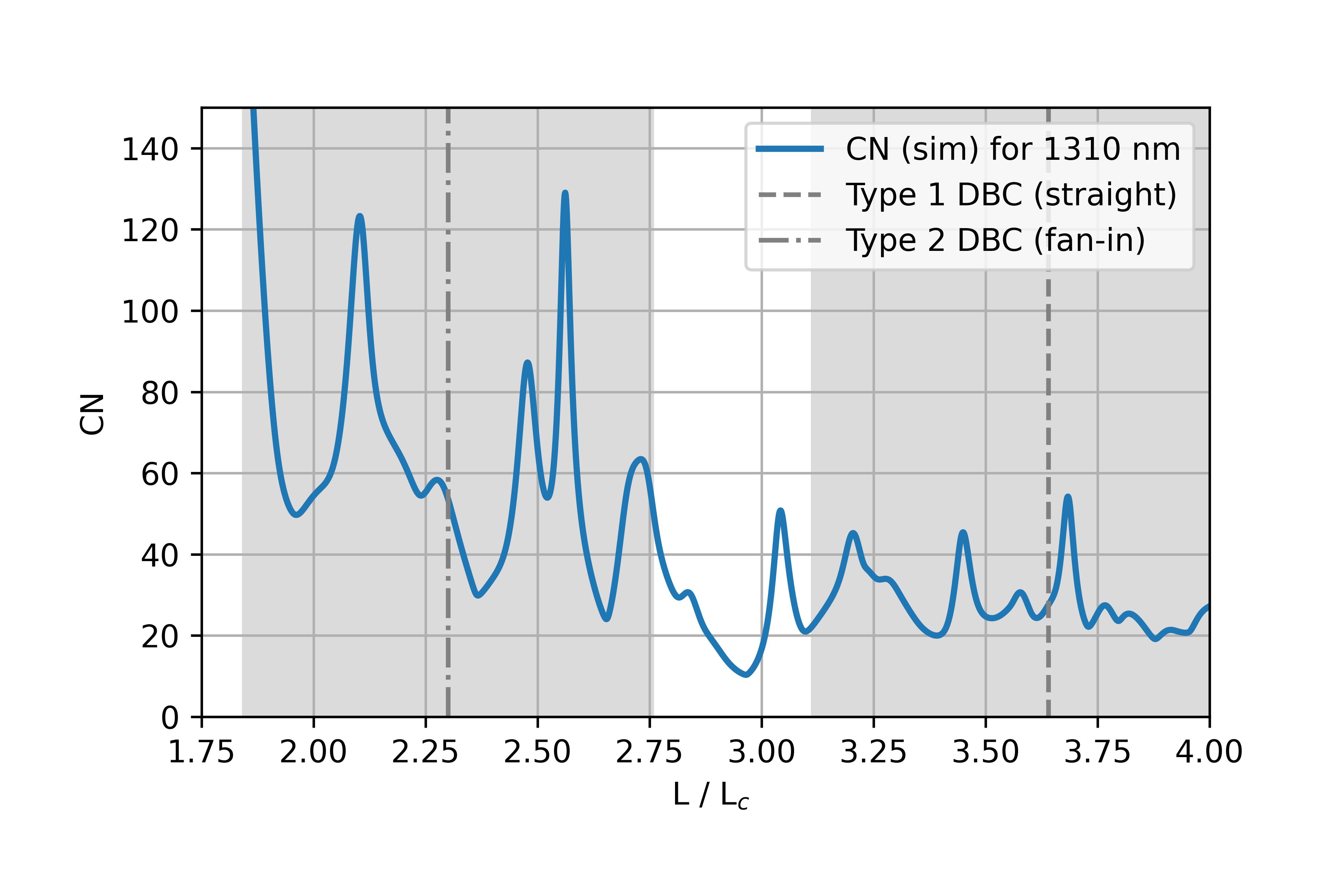}

\caption{Numerical simulation of the expected condition number CN, for simulated coupling at a wavelength of 1310~nm. The dashed line indicates the average $L/L_C$ of the type 1 (straight, $L/L_C = 3.64$) and the dashed-dotted line of the type 1 (fan-in, $L/L_C = 2.3$) DBC devices that are presented here. The gray shaded area represents the standard deviation of $L/L_C$, which is calculated from the statistical measurement uncertainty for the coupling constant $\kappa$. Note that the y-axis is cropped to only show the relevant range. For $L/L_C < 1.8$, the CN rapidly increases.}
\label{fig:CN_sim_type1and2}
\end{figure}

As an estimate of the stability of the V2PM and the suitability of using its inverse, the P2VM, to extract the visibilities, is the condition number CN, see also~\cite{Minardi:2012}. It is a measure of how errors at the input translate to errors at the output of a linear system. Here, errors in the measurement of the intensity at each output WG would translate into errors of the coherence vector and thus the visibility, with a scaling (or amplification) that can be described by the CN. Lower values for the CN describe a more stable matrix, with an ideal case of CN = 1. For the experimentally obtained V2PM in this work, the CN is computed using Python. 
Prior to fabrication, the CN was computed for simulated DBCs of different length ratios $L/L_C$ using Matlab (based on scripts by S. Minardi). This gives us expected values between 20 and 60 for the type 1 (straight) DBC device and between 20 and 130 for the type 2 (fan-in) DBC device, see Fig.~\ref{fig:CN_sim_type1and2}.

\subsection{Extracting visibilities using the P2VM}
\label{Vissectionmethod}
With the V2PM filled, it can be verified by applying the analysis to the same or different data of the same kind (in terms of setup, step size, wavelength). If required, the reduced data can be smoothed using a Savitzky-Golay-Filter~\cite{Savitzky:1964}.  The coherence vector $\vec{J}$ is obtained by performing a matrix multiplication of the normalized data with the P2VM for each baseline. From the resulting coherence vector, the normalized visibility amplitude (the Michelson visibility) and the phase are extracted for each recorded sample of each baseline using the following two equations~\cite{Diener:17,Saviauk:13}:

\begin{equation}
    V_{ij}=\sqrt{\frac{(\mathfrak{Re}\Gamma_{ij})^2+(\mathfrak{Im}\Gamma_{ij})^2}{\Gamma_{ii}\Gamma_{jj}}} \qquad(i\neq j)
\label{eq:vis}
\end{equation}
\begin{equation}
    \phi_{ij}=\arctan\frac{\mathfrak{Im}\Gamma_{ij}}{\mathfrak{Re}\Gamma_{ij}} \qquad(i\neq j)
\end{equation}

Note that the retrieved visibility presented here is an estimate of the calibration source visibility, without information about the instrumental contrast, which is self-calibrated~\cite{Tatulli:2007}. Loss in instrumental contrast can be due to a combination of effects, both in the optical system of the test bench and in the DBC chip, e.g. polarization mismatch, beam imbalance or long-range stress-induced birefringence during manufacturing~\cite{Diener:17}. The instrumental contrast can be calculated from the V2PM using Eq.~8 in~\cite{Diener:17}, and we list an average result in Tab.~\ref{tab:DBCABCD}. Instead of the instrumental contrast, we use the CN of the V2PM as an estimate of the DBC's stability. Throughout the paper, unless otherwise specified, the term visibility describes the extracted visibility of our laboratory calibration source, with the visibility amplitude as defined in Eq.~\ref{eq:vis}, and not the instrumental contrast.

\section{Results of the DBC characterization}
\label{results}
The characterization is performed for each of the selected DBCs at different wavelengths in the J- and H-band, the wavelength coverage is limited by the tunability of the laser (1280 - 1380 nm for the J-band). Since the type 1 (straight) DBC device was a proof-of-concept design, limited data was recorded that mainly aimed to support the type 2 design. A wider range of data is available for the type 2 device, which will be the focus of the result section. Here, several monochromatic measurement runs have been performed, of which the following are presented: The measurement run labeled M1(J) in this paper includes characterization from 1280 to 1380 nm in steps of 2 nm. M2(J) and M3(J) include measurements from 1332 to 1380 nm in steps of 4 nm. Also analyzed is a small dataset for the H-band, with measurements at 1520 nm, 1550 nm, and 1580 nm. The first part, section~\ref{resultsCN}, describes the V2PM in terms of its stability using the CN. The second part, section~\ref{visresults}, presents the extracted visibility using the DBC, with an estimate of the stability in section~\ref{StabilitySec}. Results of the broadband measurements are shown in section~\ref{BroadbandSec}. 

\subsection{Wavelength-dependent V2PM and CN}
\label{resultsCN}
\begin{figure}
\centering \includegraphics[width=0.9 \textwidth]{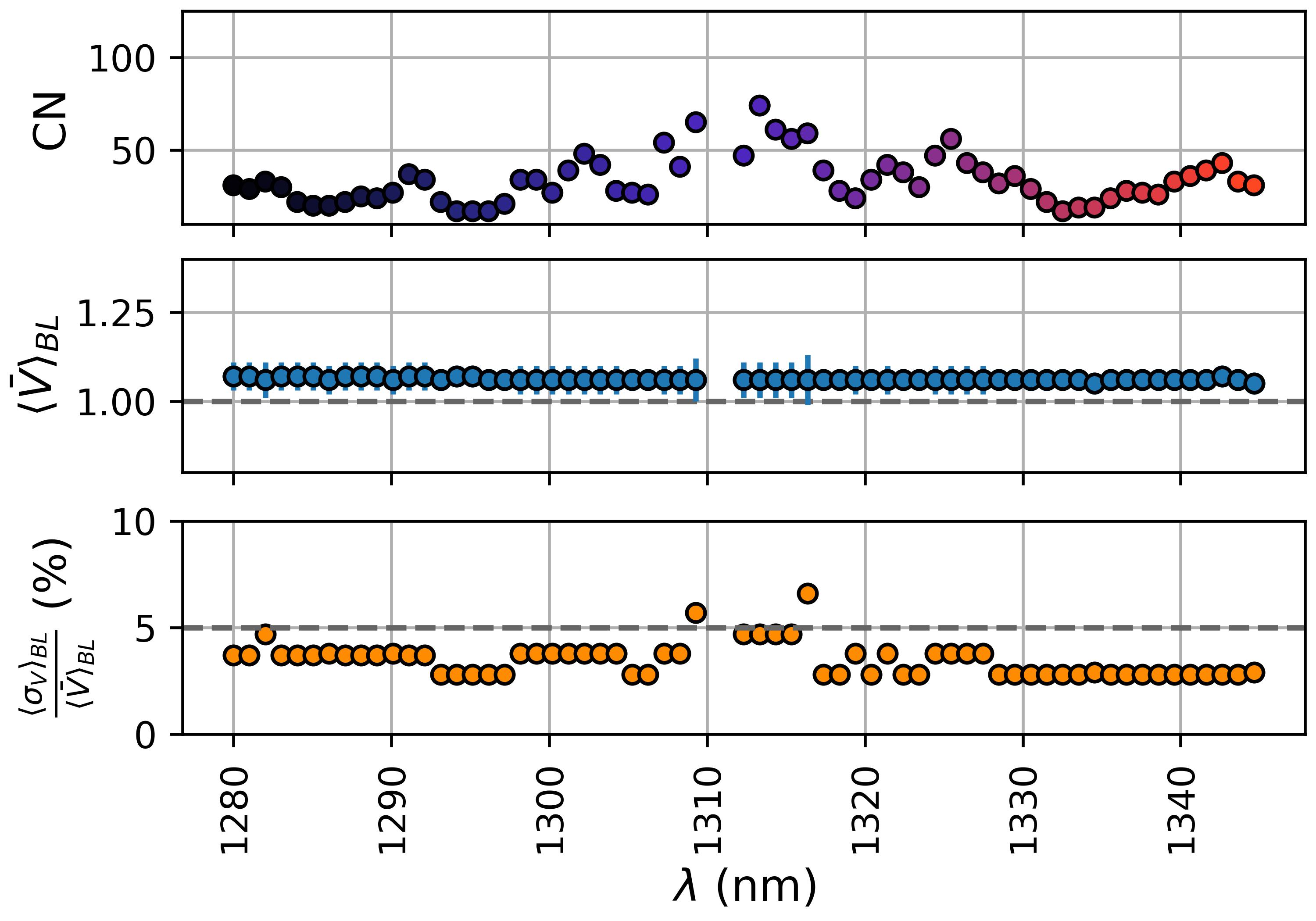}
\centering \includegraphics[width=0.9 \textwidth]{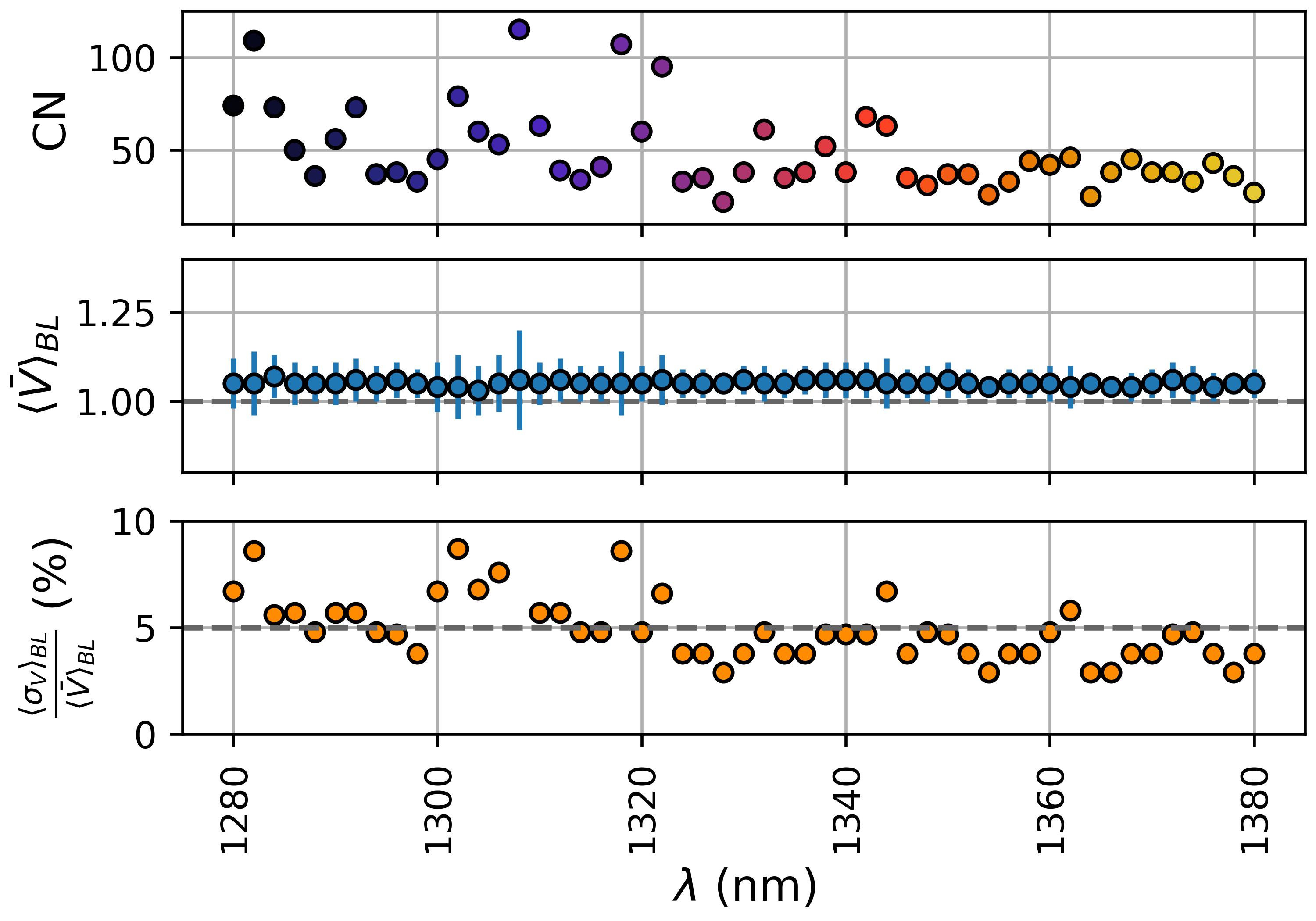}

\caption{Results for the two types of 6T-DBC devices. Top: Type 1 (straight) 6T-DBC device. Shown are the condition number CN as well as averages over all baselines of the mean visibilities \Visav with uncertainties \Sigav and relative precision \Errav at different wavelengths in the J-band (1280 - 1345~nm with $\approx$1~nm steps). The data at 1310 - 1311~nm had to be excluded due to misalignment during the delay line movement. Bottom: Type 2 (fan-in) 6T-DBC device. Shown are CN, \Visav and \Errav for 1280 - 1380~nm (in 2~nm steps). A change of characteristics and performance with wavelength is expected due to wavelength-dependent coupling.}
\label{fig:Analysis_results_Jbandtype1and2}
\end{figure}

For all measurement runs and at each wavelength, the corresponding V2PM is obtained. Since the CN is an indicator for device performance, the CN is plotted as a function of wavelength in Fig.~\ref{fig:Analysis_results_Jbandtype1and2} for the type 1 (straight) and the type 2 (fan-in) DBC device. Both devices show troughs and peaks similar to what was obtained from numerical simulation, see~Fig.~\ref{fig:CN_sim_type1and2}, and their experimental CN values lie around the values simulated for 1310~nm. The CN values for the type 1 DBC are between 17 and 74 for 1280 - 1345~nm. The CN has local minima around 1295~nm and 1333~nm (where the best performance can be expected), with a CN maximum around 1315~nm. The type 2 (fan-in) device has experimental CN values between 22 and 115, with the minimum CN at 1328~nm. With a lower coupling constant $\kappa$, the type 2 DBC performance was expected to be different. The type~2 DBC seems to generally perform better with longer wavelengths: From 1346~nm to 1380~nm, the CN is below 50, with several values close to the minimum. This trend can be explained with wavelength-dependent coupling: longer wavelengths shift the results towards higher $L/L_C$, thus move towards the lower CN region in Fig.~\ref{fig:CN_sim_type1and2}.

\subsection{Extracted visibility} 
\label{visresults}
\begin{figure}
\centering\includegraphics[width=0.9 \textwidth]{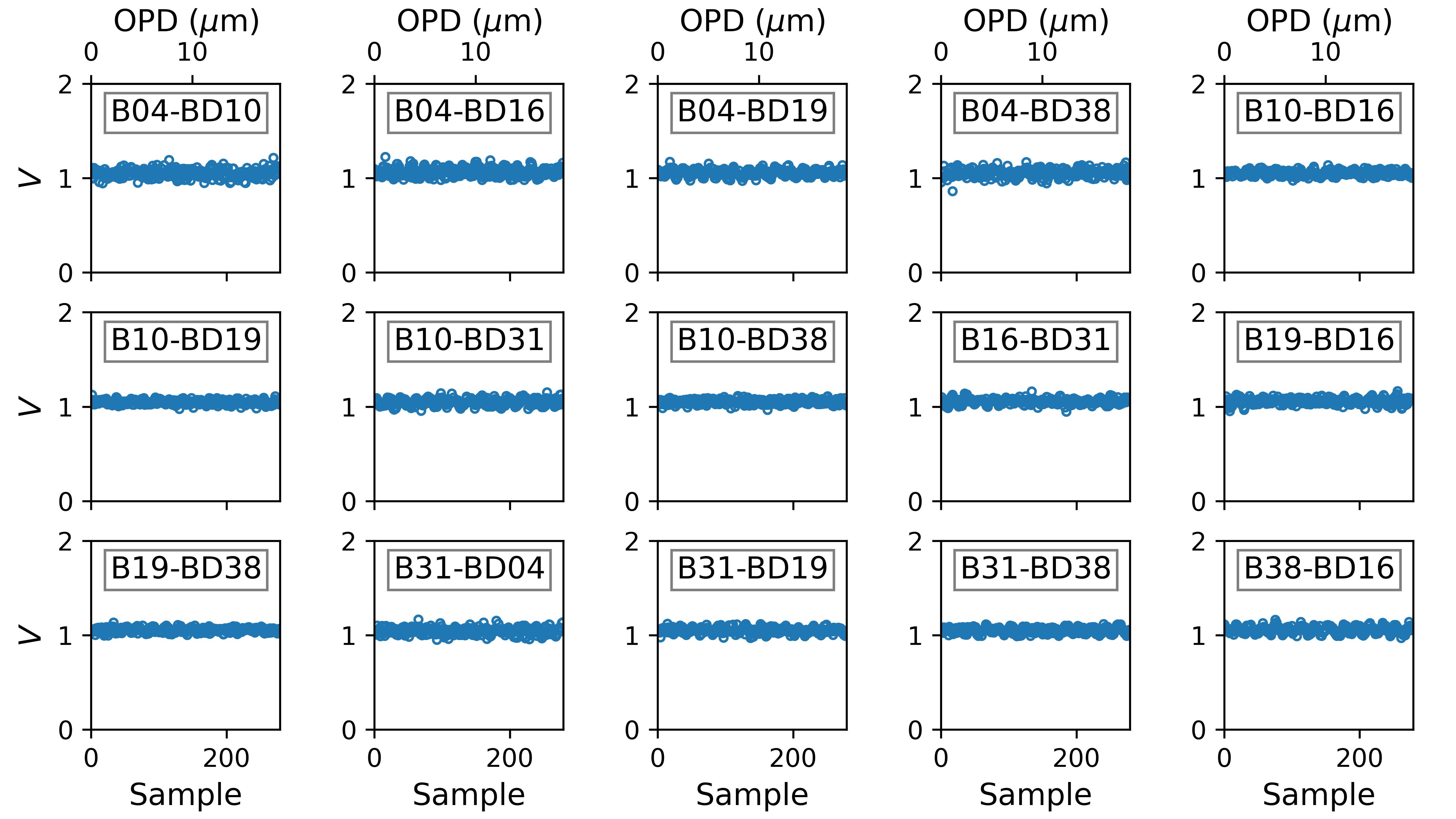}
\centering\includegraphics[width=0.9 \textwidth]{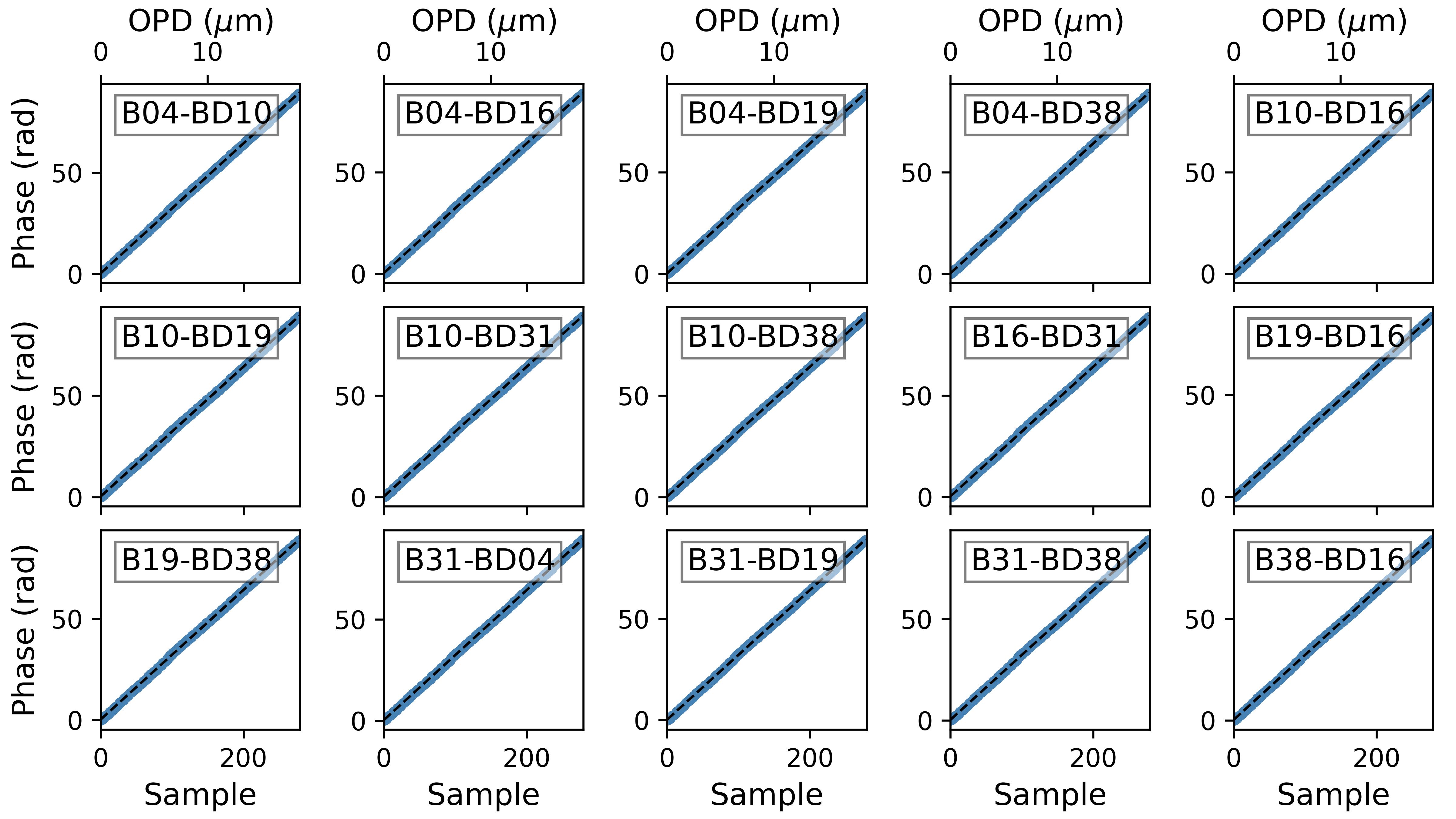}
\caption{Visibility amplitude and unwrapped phase derived from the complex visibility for 15 baselines at 1328~nm. The expected phase calculated from the frequency fit and the delay line step is indicated with a dashed line.}
\label{fig:1328amplitudeandphase}
\end{figure}

The data analysis follows the method described in section~\ref{Vissectionmethod} to extract the visibility from the data. In the first instance, to avoid stability issues, the data used for analysis is the same that was used to obtain the V2PM. No smoothing was applied to the monochromatic data presented here. Examples for extracted amplitude and phase are shown for the type 2 (fan-in) DBC device at 1328~nm in Fig.~\ref{fig:1328amplitudeandphase}. The results are shown for all 15 baselines. A visibility amplitude of 1 is expected for our point source. Per baseline, the mean visibility amplitude and its standard deviation is calculated over all samples, $\bar{V} \pm \sigma_V$, which is also used to define the relative precision \RP. For easier comparison between datasets, the average over all 15 baseline is calculated, denoted \Visav. In this example, we experimentally extract a value of \Visav = 1.05 with an averaged (over all baselines) relative precision, \Errav, of $2.9\%$. Since the light is monochromatic, the phase should evolve linearly with delay line movement, which is calculated from the fitted wavelength and marked in the plot. 

A summary of the averaged extracted visibility and relative precision for type 1 (straight) and type 2 (fan-in) DBCs in the J-band is included in Fig.~\ref{fig:Analysis_results_Jbandtype1and2} below the corresponding CN. Visually, the relative precision \Errav follows the shape of the corresponding CN, with lower relative precision values achieved for lower CN values. This means that for the type 2 (fan-in) device, the performance improves for longer wavelengths, at least in the J-band. For both types of devices, the lowest (average) relative precision achieved is just below $3\%$. The average visibility amplitude is close to the expected value of 1. Table~\ref{tab:VisJH} summarizes important performance results. 

\begin{table}[htb]
    \centering
    \begin{tabular}{l|l|l|l|l|l}
                \textbf{Design} & \textbf{Example} & \textbf{$\lambda$ (nm)} & \textbf{CN} & \textbf{Av. Visibility} & \textbf{Av. Precision}\\
                  &  & &  & \Visav $\pm$ \Sigav & \Errav\\
         \hline
       Type 1 (straight)  &Best (J-band) & 1294 & 17 & $1.07 \pm 0.03$ & 2.8\% \\
         \hline
    
        Type 2 (fan-in) &Best (J-band) &1328 & $22$ & $1.05 \pm 0.03$ & 2.9\% \\
         & Mono (J-band) & 1340 & 38 &  $1.06 \pm 0.05$& 4.7\%  \\
         & Mono (J-band) & 1350 & 37 &  $1.06 \pm 0.05$& 4.7\%  \\
         & Mono (J-band) & 1380 & 27 &  $1.05 \pm 0.04$& 3.8\%  \\
         &Broad (J-band) & 1350 & 32 &  $1.03 \pm 0.05$& 4.9\%  \\
         &Best (H-band) & 1520 & 34 & $1.04\pm 0.05$ & 4.8\% \\
         & Mono (H-band) & 1550 & 23 & $1.04\pm 0.07$ & 6.7\% \\
         &Broad (H-band) & 1550 & 26 & $1.04 \pm 0.05$ & 4.8\%  \\
    \end{tabular}
    \caption{Overview of CN, relative precision, and visibility amplitudes and their sample standard deviations for J-band and H-band measurements. For both, the extracted mean visibility amplitude and its standard deviation, the average value over all 15 baselines are given. From these average values, the average relative precision is calculated in the last column. Chosen are wavelengths where the relative precision is lowest ('Best') and some that correspond to the data shown in the Figures within this paper. Dataset M1(J) is used for type 2 monochromatic J-band data. Trimmed data containing $\approx 10$~fringes around the mean zero optical path difference is used for the broadband characterization.}
    \label{tab:VisJH}
\end{table}

\begin{figure}
\centering\includegraphics[height=0.85 \textheight]{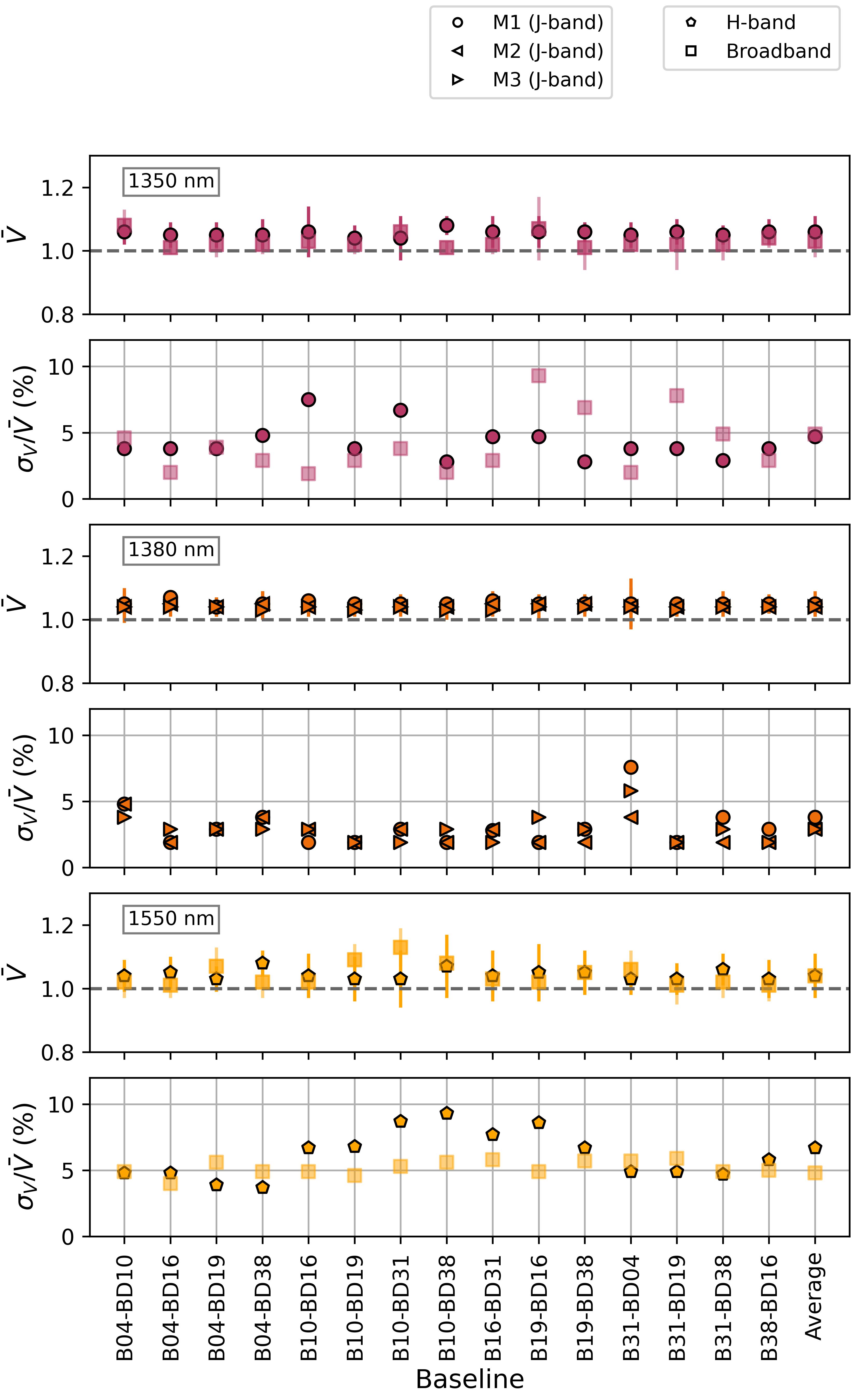}

\caption{Mean visibilities $\bar{V}$ with uncertainties $\sigma_V$ as well as relative precision for 15~baselines (and the average over the baselines) at different wavelengths. The symbols refer to different measurement runs. Note that the squares in 1350~nm and 1550~nm mark broadband measurements (using the central $\approx10$~fringes) for which the visibility amplitude extraction is affected by dispersion effects. }
\label{fig:Vissummary}
\end{figure}

Figure~\ref{fig:Vissummary} presents the mean visibility amplitude $\bar{V}$ and relative precision for each individual baseline for the type 2 design. Two wavelengths in the J-band (1350~nm and 1380~nm) have been selected for their reasonably low CN, with additional broadband data available at 1350~nm and multiple datasets at 1380~nm. From the H-band datasets, 1550~nm was selected since broadband measurements have been performed at the same central wavelength. The monochromatic data from M1(J), M2(J), and M3(J) at 1380~nm show similar variations between baselines at each wavelength (e.g. slightly higher relative precision values for baseline 31-04). Differences in precision between the baselines at a given wavelength are influenced by the signal-to-noise ratio (SNR) of the individual WG outputs. No baseline has a particularly low performance at all wavelengths, the variation is rather wavelength-dependent as light spreads differently through the array. 

\begin{figure}
\centering\includegraphics[width=0.75 \textwidth]{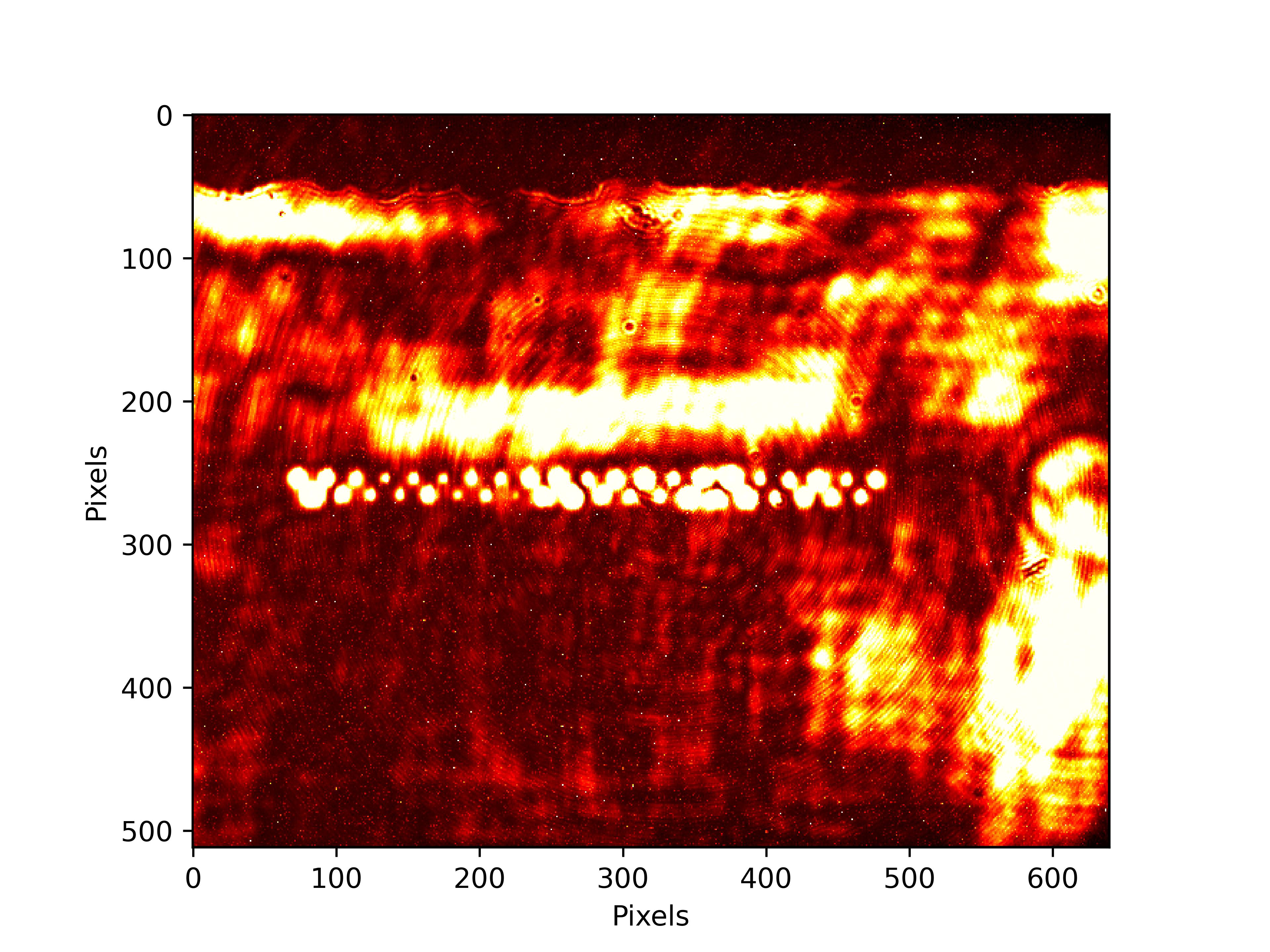}
\caption{Image of the output facet of the chip, showing the 41 output WGs of a DBC and background scatter from uncoupled light at the input sites. Here, the integration time is increased to 517~ms ($\sim 100~\times$ higher than what is used for measurements), which saturates the output WGs but visualizes the scatter and its patterns. One edge of the chip can be seen near the top. The scatter varies depending on the input site and might influence the visibility extraction.}
\label{fig:BKGscatter}
\end{figure}

For the measurements presented here, we see a systematic effect: In addition to the statistical precision of $\bar{V}$ and \Visav, it appears that the visibility amplitude is overestimated by around $0.05$, with extracted values for \Visav between $1.0$ and $1.1$, see e.g. table~\ref{tab:VisJH}. Without a full error model, this systematic is not fully understood yet and will be subject to further studies. We suspect, however, that background scatter from imperfect coupling is responsible for this shift in visibility. Interferometry is generally sensitive to photometric imbalances, with errors in the photometries even amplified in a V2PM. We do not have photometric taps in the devices presented here, thus photometric measurements for the normalization were taken subsequently rather than simultaneously. The background differs between the photometric measurement (one beam is injected) and the interferometry measurement (two beams are injected), with spatially varying background due to interference. Background scatter has been shown to affect measurements, in particular of the closure phase for the pupil remapper in~\cite{Norris:2014}. Similarly to one of their examples, the input and output WGs are not significantly displaced from each other in our 6T-DBCs. Accordingly, we can visualize the background scattering and emphasize it with an increased camera integration time, as shown in Fig.~\ref{fig:BKGscatter}. Future DBC designs should consider not only photometric tap WGs but also a displacement between input and output WGs, such as the "side-step" in~\cite{Norris:2014}. Devices might also be connectorized with fibers to reduce scatter.

\subsubsection{Extended application: lower number of inputs}
Interestingly, a six-input DBC can also be used for lower numbers of telescopes without losing the light -- provided that light from all inputs is sufficiently spreading through the array. With our pairwise characterization, we can imitate this scenario by considering only all data for four inputs (6 baselines) from an existing dataset and discarding the remaining raw data. For the 1380~nm dataset of the M1(J) measurement, we try two configurations: Removing all data for the outer inputs (4 and 38) results in a CN of 8, \Visav = 1.05, and an average relative precision \Errav = 1.9\%. Keeping the outer inputs and instead removing data for inputs 10 and 19 (so that the four inputs are roughly evenly spaced within the WG array) gives a similar result, with CN = 6 , \Visav = 1.05, and an average relative precision \Errav = 1.9\%. This means that DBCs for larger number of telescopes have the potential to serve as beam combiners for interferometry with less inputs without discarding light for unrequired overlap, e.g. in the context of test benches where the number of available telescopes for a particular instrument might vary. 

\subsubsection{Static visibility measurement}
One of the advantages of the DBC is that after the calibration, fringe scanning is not required for visibility extraction. Instead, the interferometric measurement can be static and the visibility can --in principle-- be obtained from a single shot. For one example, at 1380~nm, static interferometry data was recorded with the type 2 (fan-in) device, where both inputs are addressed, but the delay line was motionless. A calibration measurement was taken shortly before the static measurement. The number of recorded frames was kept the same between calibration and static interferometry to allow better comparison of the statistical uncertainty. For the best recorded case, the extracted \Visav is 1.04 and the average relative precision \Errav = $1.0 \%$, with $\sigma_V/\bar{V} = 0.9 - 2.0 \%$ for the different baselines. These values are lower than what has previously been measured when scanning the delay line, which indicates that delay line motion affects the relative precision. The remaining uncertainties in this static measurement likely stem from a combination of changes of the light intensity at the inputs, laser stability, vibrations, and camera noise. Photometries (not taken simultaneously) show variations from typically 1\% - 3\%, but up to 30\%. For this measurement, the time between calibration and visibility measurement was deliberately kept short. The long-term stability of the calibration, which strongly depends on the stability of the experimental setup, will contribute to the uncertainty when measurements are further apart. 

\subsection{Stability of the calibration process} 
\label{StabilitySec}
The three measurement runs M1(J), M2(J), and M3(J) have been recorded on different days, 02.~March~2023, 06.~March~2023 and 07.~March~2023, respectively. According to Fig.~\ref{fig:Vissummary}, their performance is in general agreement at 1380~nm. While this is a good starting point for a stable system, a closer look reveals that the lifetime of the V2PM calibration might be limited, likely due to the experimental setup. Since the monochromatic light source is not actively stabilized, the calibration wavelength can both drift (e.g. due to temperature) or jump (due to mode-hops) over time. This was recorded to be less than 1~nm over 30 minutes. Longer timescales have not been monitored yet. However, DBC characterization measurements at the same wavelength that are performed on separate days --and with the laser switched off or shifted in-between-- slightly differ. The (fitted) wavelength of one measurement does not match the (fitted) wavelength of another, e.g. for a set wavelength of 1380~nm, the interference fits convert to wavelengths of 1382~nm and 1376~nm for M2(J) and M3(J), respectively. While the accuracy of the fitted values is left to be determined, the wavelength difference clearly shows in the data when the V2PM from one measurement is applied to the data of another: the measured phase linearly diverges from the expected value and the visibility amplitude oscillates. This wavelength mismatch increases the standard deviation of the extracted visibility, e.g. a relative precision of 4.9\% instead of 2.9\% is measured when the V2PM of M2(J) is used to extract the visibility from M3(J) data. Actively frequency-stabilized laser sources might improve the stability and repeatability of any monochromatic calibration. For broadband operation, this will not be important. Here, other drifts are still relevant, such as light intensity and coupling stability. To address this, better temperature control of the setup, motorized mirrors that couple light into the inputs with higher repeatability, or an additional automated optimization step may be required. 

\subsection{Broadband light source}
\label{BroadbandSec}
To assess the suitability of the DBC for star light, a small dataset with broadband light was taken. The chromatic behavior of DBCs has been the result of previous studies~\cite{Nayak:2020}, where the bandwidth dependence of the V2PM formalism, the CN, and the subsequent accuracy of visibility extraction for a DBC has been shown. The studies conclude that DBCs should be operated with light with bandwidths < 100~nm~\cite{Nayak:2020}, but we chose even narrower bandwidths to further limit chromatic effects. For the H-band, the Amonics ALS-CL-15 light source was used, with a 40 nm bandwidth filter centered at 1550 nm $\pm 8$~nm (Thorlabs FB1550-40). For the J-band, the LEUKOS ELECTRO 250 IR was used, with an added 12 nm bandwidth filter centered at 1350~nm $\pm 2.4$~nm (Thorlabs FB1350-12). To observe fringes, the optical paths in the Michelson setup have to be matched to within the coherence length. Prior to the measurement, the distance from the two mirrors to the bulk optics beam splitter is matched by manually adjusting the delay line position. For the fine adjustment, the delay line is scanned while the DBC device is live recorded on the camera, until flickering of the light at the outputs indicates interference. Since the same two mirrors are used to couple into the different input sites by changing the mirror angle, the delay line position for fine-adjustment of the OPD depends on the input sites. 

\begin{figure}
\centering\includegraphics[width=0.9 \textwidth]{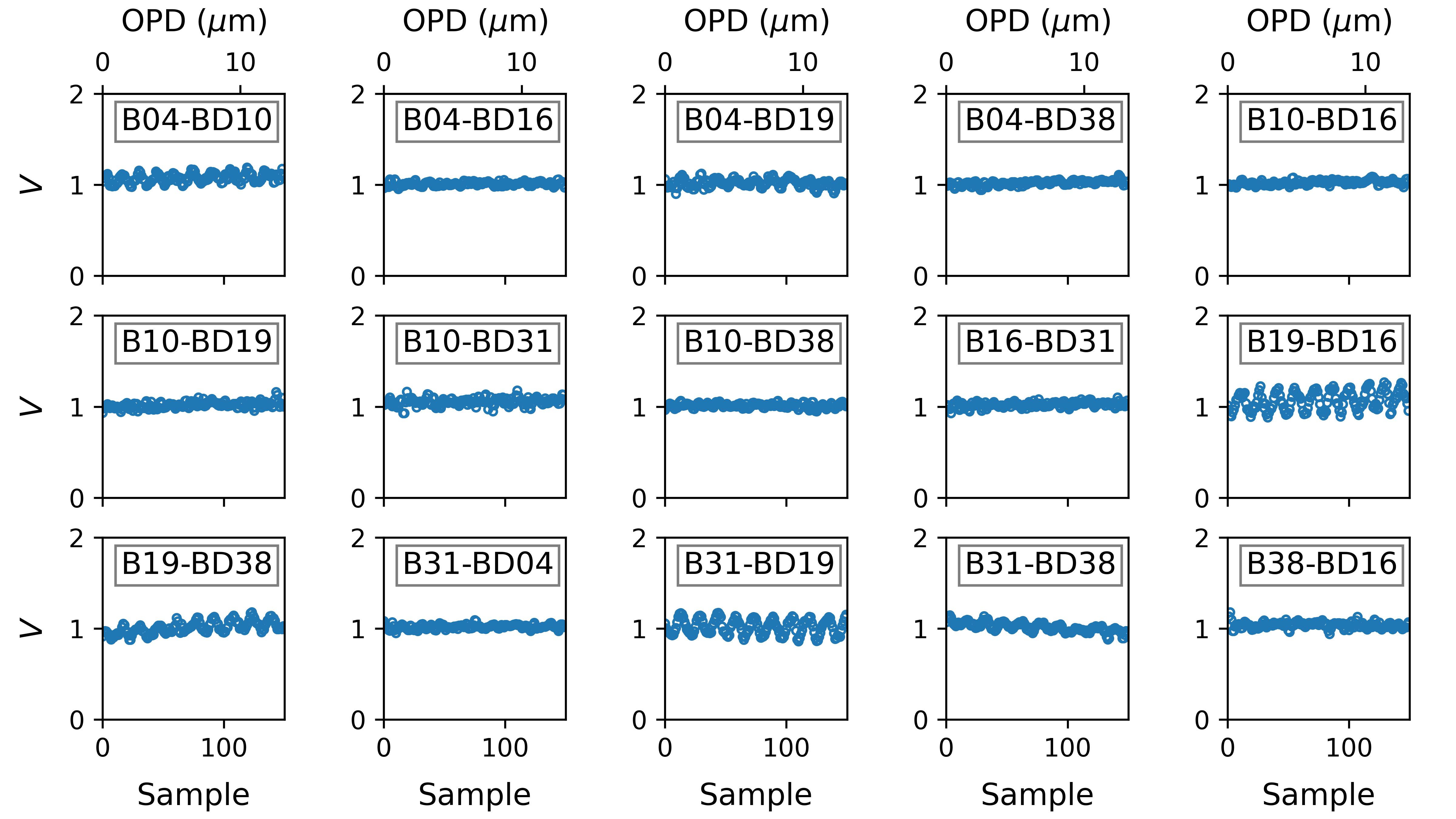}
\centering\includegraphics[width=0.9 \textwidth]{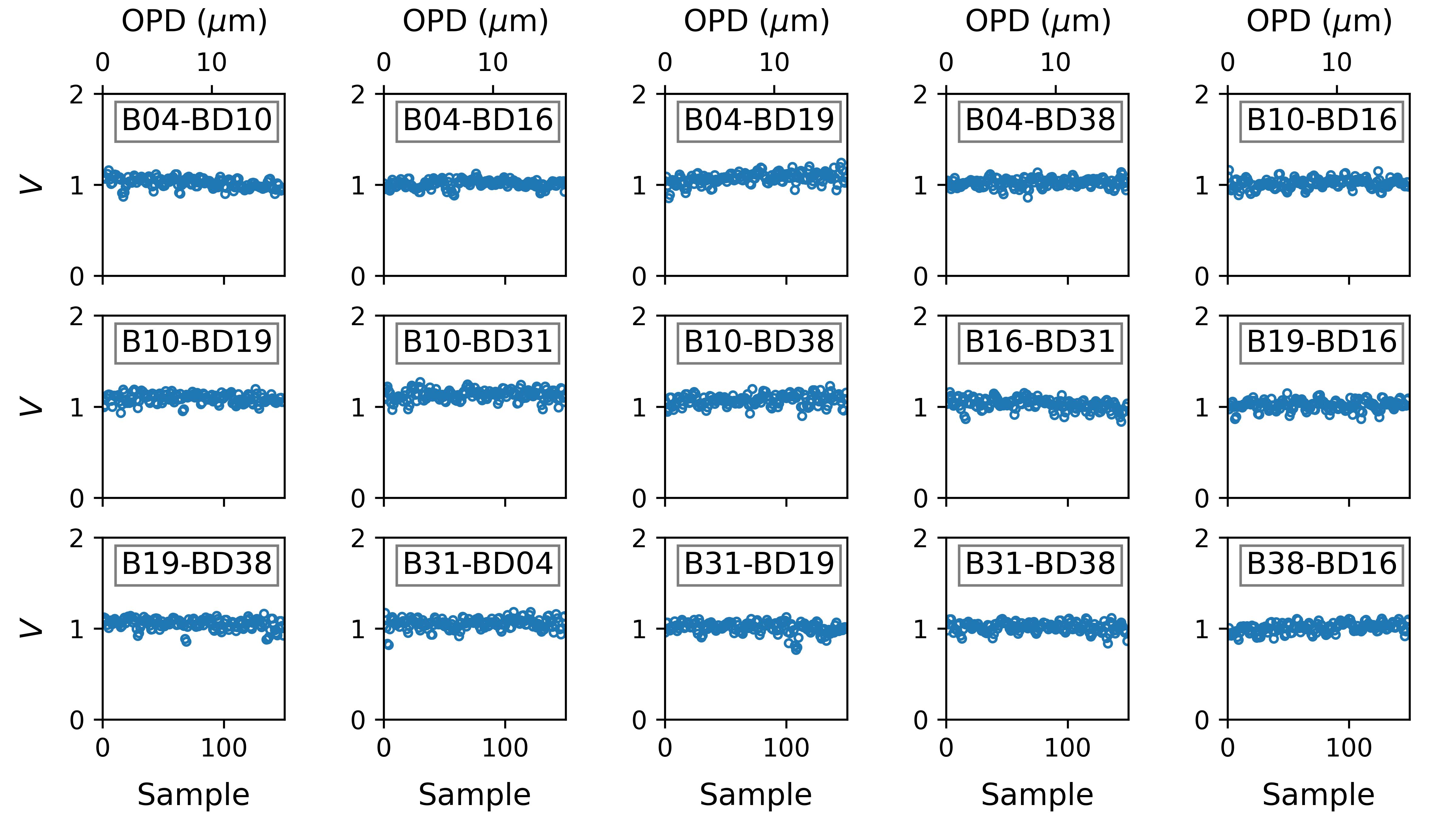}
\caption{Broadband measurements: Extracted visibility amplitude for the J-band ($\lambda = 1350$~nm, $\Delta \lambda = 12$~nm, top) and in the H-band ($\lambda = 1550$~nm, $\Delta \lambda = 40$~nm, bottom). The x-axes show the sample numbers (bottom) and the OPD caused by the respective delay line movement (top) (therefore relative to the first data point shown, with the absolute ZPD in the center). }
\label{fig:BroadAmp}
\end{figure}

Due to material properties as well as wavelength dependent coupling (see e.g.~\cite{Nayak:2020}), the recorded fringes show dispersion effects in the form of e.g. skewed envelope functions. This effect is stronger for the H-band measurements, where a 40~nm bandwidth was used, which is the main reason for choosing a different, narrower width for the J-band measurements. To avoid these effects in future, dispersive elements can be added at the output of the DBC device. 

For the calibration of the broadband V2PM, an additional step is required to process the data, which contains around ten times more frames than monochromatic data. A maximum search is performed on the fringe signal at each output WG to identify the maximum amplitude and thus (ideally) the position of zero optical path difference (ZPD). From the maximum positions for all WGs, the median is taken to obtain one ZPD position per baseline. The data that is used for the matrix calibration is cut to contain roughly the central 10 fringes around the ZPD, corresponding to around 150 samples. For the verification step, where the data is multiplied with the P2VM to obtain the visibility, the length of the data is also limited to the central fringes, but could be increased if desired, e.g. to see the envelope shape. Before multiplication with the P2VM, the data is smoothed using the Savitzky-Golay-Filter (polynomial of order 4 with filter window size = 9 for J-band data and filter window size = 7 for H-band data). 

The J-band characterization with the broadband source results in a CN of 32, which is comparable to the monochromatic CN at the central wavelength of 1350~nm (CN=37). For the H-band characterization, the CN is 26, which is close to the value for the monochromatic measurement at 1550 nm (CN=23). For broadband data, an increase of the CN is generally expected compared to monochromatic measurements due to the chromaticity of the V2PM, see~\cite{Nayak:2020}. Here, the chosen filter bandwidths seem to enable comparable results. The average visibility amplitudes and corresponding relative precision are given in Tab.~\ref{tab:VisJH} for the analysis of the central 10 fringes, which match the monochromatic results. Outside the central fringes, the visibility amplitude drops off due to the Sinc envelope function. As for the monochromatic data, wavelength- and baseline-dependent performance differences can be seen in Fig.~\ref{fig:Vissummary}. 

The visibility amplitudes for the J- and H-band are shown in Fig.~\ref{fig:BroadAmp}. Here, the J-band (broadband) visibility amplitudes have visible oscillations. For monochromatic data, this can be an indicator that the frequency fit during the matrix calibration is not correct. A visual inspection, however, shows good agreement. Oscillations have actually been predicted for broadband operation in~\cite{Nayak:2020} from simulations. Similar oscillations were also found for experimental data in~\cite{Nayak:21}. These oscillations are due to the inherent wavelength dependence of the V2PM and appear because of the current V2PM calibration method. Developing an improved V2PM calibration procedure might help to surpass these oscillation effects in future. They do affect the performance parameters, since the baselines with the strongest oscillations show the lowest relative precision (highest percentage values), see Fig.~\ref{fig:Vissummary}. The H-band does not show oscillations, but the effect might have been disguised, e.g. by noise. 

An aspect that could be investigated further (maybe in relation to above effect) is the maximum fringe position. When considering the individual fringes at the 41~output WGs for a wider delay line range (not shown), we see that the location of the white light envelope maximum differs between the fringes at the individual WGs. Since we take the same ZPD position for each baseline (the median of the ZPDs of all WGs), the individual WG fringes can have their maxima offset to that position, leading to contributions to the V2PM characterization at non-ideal positions. Using the median ZPD is our approach of a compromise and dismissing outliers (due to strong dispersion effects or changes in the background that might provide a false maximum). Nevertheless, for most baselines the WG ZPDs have large enough shifts relative to each other that some fringes have a zero-crossing where others have their maximum. With the resulting low signal and/or phase-shifts in a substantial number of WGs, the calibration might be non-ideal. 

We believe that the dispersion causes unknown phase shifts from a common relative reference that affect our retrieved results. Presently, the V2PM is limited to quasi-monochromatic sources ($\Delta \lambda \ll \lambda$). While it might be possible to find a more robust V2PM that takes into account dispersion effects for broadband light measurements, spectro-interferometry approaches are a more promising path.

\section{Summary and conclusions}

We have developed the first integrated six-telescope beam combiner for stellar interferometry in the J-band. We have manufactured the 3D structures of this DBC using the technique of ultrafast laser inscription (ULI), which enables flexible and low-cost prototyping. With six inputs, this DBC has the largest number of beam combinations of its kind. The component was written in borosilicate glass, with six inputs that are evenly spaced by 127~$\mu$m, a reformatting region, and the DBC region with 41 WGs in a 3-dimensional array over two layers. We measured transmission of 56\% and detected no polarization dependence. The beam combiner was characterized at near-infrared wavelengths using a tunable laser centered around the wavelength of 1330~nm corresponding to the astronomical J band as well as a laser in the H-band centered around 1550~nm. Additionally, first measurements with broadband light sources in the J- and the H-band were performed (1350~nm and 1550~nm, respectively), all with an automated two-input Michelson interferometer. In addition to the characterization of the fan-in DBC design (type 2), a small set of results for a proof-of-concept device (type 1) without fan-in region have been included in this paper.  

In a first step, the V2PM was experimentally determined for different wavelengths in the J- and H-band and its stability characterized in terms of the CN. In a second step, visibilities of the calibration light source were extracted. Both types of devices show wavelength-dependent performance in accordance with numerical simulations, each with several local optima that may serve as working points. The fan-in device has several potential working wavelength regions, but generally showed better performance (i.e. more stable V2PM and mean visibilities with smaller standard deviation) at longer wavelengths (> 1325~nm). We relate this to design choices and manufacturing tolerances that resulted in better suited specifications of the device for longer wavelengths, in particular the length of the DBC array and the coupling strength. One local optimum of the type 2 (fan-in) device is at 1328~nm, with a CN of $22$ and a visibility amplitude of 1.05 with an average relative precision \Errav of $2.9\%$. At 1380~nm, we obtain a CN of $27$, a visibility amplitude on the order of 1.05, and an average relative precision \Errav of $3.8\%$. The H-band performance has been evaluated for a small set of wavelengths, e.g. at 1520~nm, where CN = 34 and \Errav = 4.8\%. For the type 1 design, the optimum is found at lower wavelengths in the J-band, with CN = 17 at 1294~nm, and a relative precision of 2.8\%. In all cases, the absolute value of the visibility amplitude is larger than the expected value of 1, a systematic effect that might be caused by background scatter~\cite{Norris:2014}, but still has to be resolved. 

Broadband measurements with 12~nm (J-band) and 40~nm (H-band) bandwidths show considerable dispersion effects, but enables retrieval of the visibility amplitudes with a relative precision of 4.9\% and 4.8\%, respectively. 

At the wavelengths of best (monochromatic) performance, the extracted visibilities have a relative precision of <3 \%. Better precision values down to $1\%$ were achieved when the visibility was extracted statically, without delay line movement. Residual noise from our experimental setup might still reduce the fringe SNR at the output WGs and limit the performance by increasing measurement uncertainty. Over some period of time, vibrations have severely affected the measurements, but the main noise sources were suppressed for the data presented here. Occasionally, individual measurements have to be discarded due to unidentified spikes and distortions. Additional efforts to identify noise sources might lead to further improvements. Changes in laser wavelength and light coupling limit the stability of the V2PM and thus the reliability of visibility extraction. Further stabilization of the setup to detangle setup and DBC performance or (on-sky) tests in a known environment are required to precisely identify performance limits of the DBC device itself and allow better comparison with existing beam combiners. For example, the fiber-based two-input MONA beam combiner of the FLUOR instrument achieved stabilities of (better than) $1\%$~\cite{Foresto:2003}, and more recently, the six-telescope image-plane combiner MIRC-X at the CHARA Array has been achieving visibility precision of $1\%$~\cite{Anugu:2020}. The DBC is comparable to other photonic components in terms of throughput, albeit with a slightly less precise visibility extraction, and we imagine that it could be tested on-sky in setups similar to existing instruments. After initial population of the V2PM, the beam combiner could be used statically, and calibrated against the variable atmosphere using observations of unresolved stars close to the target. The potential range of targets as well as expected performance strongly depend on the specific telescope facility and environment, e.g. the size of the telescopes, the availability of an adaptive optics system for wavefront correction, the beam's Strehl ratio, the detector properties, and the availability of a fringe tracking system.

In general, many more tests can be performed with DBCs, such as using spectrally dispersing components at the output to circumvent dispersion effects within the DBC, upgrade of the setup to enable characterization using simultaneous coupling into all inputs, and fiber interfacing at the inputs similar to~\cite{Benoit:21}. Photometric taps and displacement between inputs and outputs should be considered to reduce systematic effects. To achieve better visibility precision, higher SNR is required at each output. Next-generation designs where the WGs in the array are detuned might enable better light spread through the array to achieve higher SNR, see also~\cite{Jovanovic:2023} for more advanced design options and challenges. To verify the 6-telescope beam combiner in more realistic settings, simultaneous beam combination and subsequent on-sky telescope tests in the J-band will be important future steps. Parallel developments to enable larger number of inputs and to investigate practical aspects of the scalability of DBCs are foreseen.


\begin{backmatter}
\bmsection{Funding}
Bundesministerium für Bildung und Forschung (03Z22AN11), Deutsche Forschungsgemeinschaft (326946494, 506421303).


\bmsection{Acknowledgments}
We would like to thank the anonymous reviewers, whose time, effort, and comments allowed us to improve the quality of the manuscript. We would also like to thank L.~Labadie, N.~Scott, E.~Hernandez, and S.~Vje\v{s}nica for valuable input.

\bmsection{Disclosures}
The authors declare no conflicts of interest.

\bmsection{Data availability} Data underlying the results presented in this paper are not publicly available at this time but may be obtained from the authors upon reasonable request.

\end{backmatter}


\bibliography{manuscript}

\end{document}